\documentclass[preprint,byrevtex,onecolumn,12pt,pre]{revtex4}%
\usepackage{amsfonts}
\usepackage{amsmath}
\usepackage{amssymb}
\usepackage{graphicx}%
\providecommand{\U}[1]{\protect\rule{.1in}{.1in}}
\newtheorem{theorem}{Theorem}

\newtheorem{conclusion}[theorem]{Conclusion}

\newtheorem{conjecture}[theorem]{Conjecture}

\newtheorem{remark}[theorem]{Remark}

\begin{document}
\preprint{UATP/1102}
\title{Is Structural Relaxation During Vitrification the Inverse of the Glass Transition?}
\author{P. D. Gujrati}
\email{pdg@uakron.edu}
\affiliation{Department of Physics, Department of Polymer Science, The University of Akron,
Akron, OH 44325}

\begin{abstract}
We have recently applied the second law to an isolated system, consisting of a
system $\Sigma$ such as a glass surrounded by an extremely large medium
$\widetilde{\Sigma}$, to show that the instantaneous absolute temperature
$T(t),$ thermodynamic entropy $S(T_{0},t)$ and enthalpy $H(T_{0},t)$ of
$\Sigma$ decrease in any isothermal relaxation towards their respective
equilibrium values $T_{0},S_{\text{eq}}(T_{0})$ and $H_{\text{eq}}(T_{0})$
under isobaric cooling. The decrease of the thermodynamic entropy and enthalpy
during relaxation in vitrification is consistent with non-negative temperature
$T(t)$. The Gibbs statistical entropy also conforms to the above relaxation
behavior in a glass, which however is contrary to the conjecture by Gupta,
Mauro and coworkers that the glass transition and the structural relaxation
during vitrification are inverse to each other; this is then supported by
computation in which their statistical entropy $\widehat{S}(T_{0},t)$ drops
below $S_{\text{eq}}(T_{0})$ during the glass transition and then increases
towards it during isothermal relaxation. However, they do not establish that
the entropy loss during the glass transition is accompanied by a concomitant
entropy gain of the medium to maintain the second law. These authors use a
novel statistical formulation $\widehat{S}(T_{0},t)$ of entropy based on
several conjectures such as it being zero for a microstate, but do not compare
its behavior with the thermodynamic entropy $S(T_{0},t)$. The formulation is
designed to show the entropy loss. Its subsequent rise not only contradicts
our result but also implies that the glass during relaxation must have a
negative\ absolute temperature. To understand these puzzling results and the
above conjecture, we have carried out a critical evaluation of their
unconventional approach. We find that the inverse conjecture is neither
supported by their approach nor by the second law. The zero-entropy microstate
conjecture is only consistent with $\widehat{S}(T_{0},t)\equiv0$ at all
temperatures, not just at absolute zero and is found to have no scientific
merit. We show that the maximum entropy gain of the medium during the glass
transition invalidates the entropy loss conjecture. After pointing out other
misleading, confusing and highly exaggerated statements in their work, we
finally conclude\ that their unconventional statistical approach and
computational scheme are not appropriate for glasses.

\end{abstract}
\date[January 28, 2011]{}
\maketitle

\section{Introduction}

\subsection{Conventional Approach (CA) to Glass Transition}

\begin{figure}
[ptb]
\begin{center}
\includegraphics[
height=2.5322in,
width=5.2243in
]%
{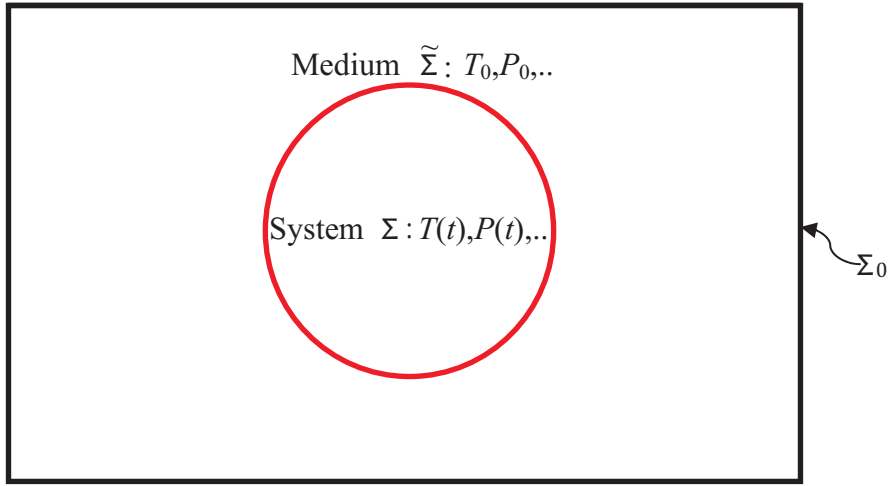}%
\caption{Schematic representation of a macroscopically large system $\Sigma$
and an extremely large medium $\widetilde{\Sigma}$ surrounding it to form an
isolated system $\Sigma_{0}$. The system is a very small part of $\Sigma_{0}$.
The medium is described by its fixed fields $T_{0},P_{0},$ etc. while the
system, when it is in internal equilibrium (see text) is characterized by
$T(t),P(t),$ etc. \ \ \ \ }%
\label{Fig_Systems}%
\end{center}
\end{figure}

In a recent paper \cite{Guj-NE-I}, we have studied a homogeneous
non-equilibrium system $\Sigma$ surrounded by an extremely large medium
$\widetilde{\Sigma}$. The work has been extended to also cover inhomogeneous
systems and internal variables \cite{Guj-NE-II}. The combined system
$\Sigma_{0}$\ forms an isolated system; see Fig. \ref{Fig_Systems}. We apply
the \emph{second law} to the isolated system to describe the behavior of the
non-equilibrium system $\Sigma$. According to the second law \cite{note1}%
\begin{equation}
\frac{dS_{0}(t)}{dt}=\frac{dS(t)}{dt}+\frac{d\widetilde{S}(t)}{dt}%
\geq0,\label{Second_Law}%
\end{equation}
where $S_{0}(t),S(t)$ and $\widetilde{S}(t)$\ denote the entropy of
$\Sigma_{0},\Sigma$ and $\widetilde{\Sigma}$, respectively, at time $t$.\ In
this work, quantities pertaining to $\Sigma_{0}$ have the suffix $0$,\ the
system $\Sigma$ have no suffix, and $\widetilde{\Sigma}$ have a tilde. For
$\Sigma_{0}$, all of its (additive) observables, variables that can be
controlled by the observer, such as its energy $E_{0}$, volume $V_{0}$, number
of particles $N_{0}$, etc. are \emph{constant} in time. These observables also
identify the macrostate of $\Sigma_{0}$. It is clear that for some homogeneous
$\Sigma_{0}$, the variation of its instantaneous entropy $S_{0}(t)$\ cannot be
explained by the dependence of the latter on its constant observables. The
variation can only be explained by assuming the dependence of $S_{0}(t)$ on
(additive) internal variables, variables that cannot be controlled by the
observer, that vary in time as $S_{0}(t)$ approaches its maximum value. For an
inhomogeneous $\Sigma_{0}$, the variation of its entropy $S_{0}(t)$\ can be
explained by the way the inhomogeneity disappears \cite{Guj-NE-II} as it
approaches equilibrium. The inhomogeneity gives rise to induced internal
variables; see below. One may not require any additional internal variable.
Thus, if we consider $\Sigma_{0}$ to consist of $\Sigma$ and $\widetilde
{\Sigma}$, as we do in this work, it is no longer a homogeneous system as long
as $\Sigma$ is not in equilibrium with the medium $\widetilde{\Sigma}$. Thus,
it is possible to consider\ $\Sigma_{0}$ without any internal variable, as was
the case studied in \cite{Guj-NE-I}. We discover that the \emph{instantaneous}
values of its \emph{fields} (temperature $T(t)$, pressure $P(t)$, etc.) are in
general different from those of the medium ($T_{0},P_{0},$ etc.); see also
Bouchbinder and Langer \cite{Langer}. But the most surprising result of the
mathematical analysis is that the instantaneous conjugate variables, the
entropy $S(t),$ the volume $V(t)$, etc. play the role of inhomogeneity-induced
internal variables with the corresponding "affinity"
\cite{Donder,deGroot,Prigogine,Guj-NE-I,Guj-NE-II,Nemilov-Book}, respectively,
related to the deviation $T(t)-T_{0}$,$P(t)-P_{0}$, etc; see Eqs.
(\ref{Gibbs_Fundamental(t)}) and (\ref{First_Law(t)}).

\begin{figure}
[ptb]
\begin{center}
\includegraphics[
trim=1.202556in 4.010781in 2.156617in 4.018231in,
height=2.6411in,
width=4.8196in
]%
{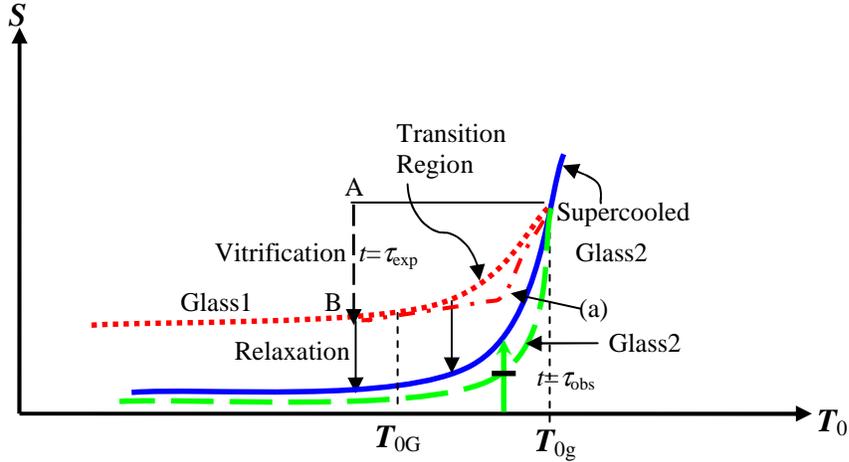}%
\caption{Schematic behavior of the entropy of the equilibrated supercooled
liquid ($S_{\text{eq}}(T_{0})$:solid curve) and two possible
glasses\ ($S(T_{0},t)$:Glass1-dotted curve, \ \ $\widehat{S}(T_{0}%
,t)$:Glass2-dashed curve) during vitrification. At $T_{0\text{g}}$, the system
falls off \ its equilibrium state, whose location depends on the rate of
cooling. Structures appear to freeze (over an extremely long period of time)
at and below $T_{0\text{G}}$; see text. The transition region between
$T_{0\text{g}}$ and $T_{0\text{G}}$ over which the liquid turns into a glass
has been exaggerated to highlight the point that the glass transition is not a
sharp point.\ For all temperatures $T_{0}<T_{0\text{g}}$, any non-equilibrium
state along the dotted and dashed curves undergoes structural relaxation in
time towards the supercooled liquid. The two vertical downward arrows show
isothermal (constant medium temperature $T_{0}$) structural relaxation at two
different temperatures in Glass1, during which the entropy $S(T_{0},t)$
decreases in time, as shown by the downward arrows. For Glass2, the entropy
$\widehat{S}(T_{0},t)$ must increase during isothermal structural relaxation.
During the relaxation, the temperature $T(t)$ of the system also decreases
towards the temperature of the medium in both glasses. The entropy of the
supercooled liquid is shown to extrapolate to zero per our assumption, but
that of Glass1 to a non-zero value and of Glass2 to zero at absolute zero. The
possibility of an ideal glass transition, which does not affect our
conclusion, will result in a singular form of the solid curve.}%
\label{Fig_entropyglass}%
\end{center}
\end{figure}

As a non-equilibrium system at fixed $T_{0},P_{0}$ of the medium strives to
come to equilibrium, it undergoes relaxation during which its instantaneous
fields $T(t),P(t)$, etc. continue to change. At the completion of relaxation,
the "affinities" vanish so that
\[
T(t)\rightarrow T_{0},P(t)\rightarrow P_{0},\ \ \ \ \text{etc.}%
\]
as expected. In an isobaric process, which is of central interest to us here,
we will assume that the system is always in mechanical equilibrium so that its
pressure $P(t)$ is always equal to $P_{0}$ at all temperatures and all times;
however, there is normally no thermal equilibrium so that the instantaneous
temperature $T(t)$ of the system is different from $T_{0}$
\cite{Guj-NE-I,Guj-NE-II,Langer}. It is found that during relaxation, the
instantaneous entropy $S(t)$ of the system continues to decrease in an
isobaric cooling experiment such as vitrification. The entropy of Glass1, see
Fig. \ref{Fig_entropyglass},%
approaches (see the downward arrows) that of the equilibrated supercooled
liquid entropy $S_{\text{eq}}$ (shown by the solid curve) from above during relaxation%

\begin{equation}
S(T_{0},P_{0},t)\overset{\text{CA}}{\rightarrow}S_{\text{eq}}^{+}(T_{0}%
,P_{0}),\label{Entropy_Behavior_CA}%
\end{equation}
where we have also exhibited the temperature of the medium; being in
equilibrium (we do not consider possible crystallization here), the
supercooled liquid cannot have any relaxation. It will be our practice to not
exhibit $T_{0},P_{0}$, unless clarity is needee, in which case we will exhibit
them. As we will mostly consider an isobaric process, we will not exhibit
$P_{0}$ in the argument. In an isobaric heating experiment, the entropy will
increase during relaxation. Such a behavior of the entropy during relaxation
will be called the \emph{conventional} behavior in this work and our approach
the conventional approach (CA) as it follows from classical thermodynamics
\cite{Donder,deGroot,Prigogine,Guj-NE-I,Guj-NE-II}. This then explains the
symbol over the arrow in Eq. (\ref{Entropy_Behavior_CA}) and the header of
this subsection. The behavior appears to be the right behavior for the entropy
in an isobaric process as the entropy must be an increasing function of the
enthalpy with the slope given by%
\begin{equation}
\left(  \frac{\partial S(t)}{\partial H(t)}\right)  _{P}\approx\frac{1}%
{T(t)}\geq0,\label{Temperature(t)}%
\end{equation}
see Eq. (\ref{dS/dH_relation}); it is well known that the enthalpy falls
(rises) during isobaric cooling (heating) \cite{Nemilov-Book}. The analysis in
our previous work \cite{Guj-NE-I,Guj-NE-II} is carried out for any
non-equilibrium system. It is therefore also applicable to glasses where one
must, in addition, make a distinction between fast and slow processes. We have
considered the issue of the fictive temperature in glasses in our work
\cite{Guj-NE-I}, where such a distinction has been made.

\subsection{Unconventional Approach (UCA) to Glass Transitions\label{Sect_UCA}%
}

In a series of papers \cite{Gupta,Mauro0,Mauro}, Gupta, Mauro and coworkers,
to be collectively denoted in short by GMc here, have developed a description
of glasses without using any internal variables or any fictive temperature.
Even the fictive temperature or pressure, which some people treat as an
internal variable, is not considered in their theoretical development,
notwithstanding the fact that they are used when GMc consider experimental
data. Their conclusion is that their statistical entropy $\widehat{S}%
(T_{0},t)$ for the glass \emph{increases} during relaxation towards
$S_{\text{eq}}$%
\begin{equation}
\widehat{S}(T_{0},t)\overset{\text{UCA}}{\rightarrow}S_{\text{eq}}^{-}(T_{0});
\label{Entropy_Behavior_UCA}%
\end{equation}
see Glass2 and the portion of the upward thick arrow in Fig.
\ref{Fig_entropyglass} above the horizontal dash; the latter is located at the
value of the Glass2 entropy. In the following, we will use $\widehat{S}$ for
the statistical entropy used by GMc to distinguish it from the thermodynamic
or the Gibbs statistical entropy $S$\ used in our approach. During the glass
transition, the entropy $\widehat{S}(t)$ \emph{falls} below $S_{\text{eq}}$ of
the supercooled liquid in this approach. Thus, they suggest that the glass
transition and relaxation are inverse processes. They also take their glass as
homogeneous, just as we have done in our earlier work \cite{Guj-NE-I}; the
difference is that their glass is identified by its temperature $T_{0}$,
pressure $P_{0}$ and its history. Thus, they make no distinction between the
instantaneous temperature of the glass and the external temperature of the
medium. As the above-mentioned "affinities" vanish, $\widehat{S}(t)\ $and
$V(t)$, etc. do not play the role of internal variables.\ Internal variables
are normally used in traditional non-equilibrium thermodynamics
\cite{Donder,deGroot,Prigogine,Guj-NE-I,Guj-NE-II}, although they are very
hard to identify, and even harder (really impossible) to control. For the
application of internal variables to glasses, we refer the reader to Nemilov
\cite{Nemilov-Book}. In the absence of any internal variable, the
approximation in Eq. (\ref{Temperature(t)}) turns into an identity
\cite{Guj-NE-I}; see also Eq. (\ref{dS/dH_relation_0}). As the behavior of the
entropy during relaxation does not obey Eq. (\ref{dS/dH_relation_0}), we will
call their approach the \emph{unconventional }approach (UCA) in this work;
this explains the symbol over the arrow in Eq. (\ref{Entropy_Behavior_UCA})
and the header of this subsection.

One can understand the drop $\Delta_{\text{GT}}\widehat{S}(t)$ in the entropy
of Glass2 during a glass transition, provided the entropy of the medium goes
up by $\Delta_{\text{GT}}\widetilde{S}(t)$ to compensate the loss and some
more so that the entropy of $\Sigma_{0}$ does not decrease%
\begin{equation}
\Delta_{\text{GT}}S_{0}(T_{0},t)=\Delta_{\text{GT}}\widehat{S}(T_{0}%
,t)+\Delta_{\text{GT}}\widetilde{S}(T_{0},t)\geq0\label{GT_Entropies}%
\end{equation}
during the glass transition. However, the authors have not discussed this
issue at all. This is not surprising as they neither include any medium in
their discussion nor do they consider an isolated system for which Eq.
(\ref{Second_Law}) holds. We will always consider $\Sigma$ as a part of the
isolated system $\Sigma_{0}$ and apply Eq. (\ref{Second_Law}) to the latter.
This will then allow us to evaluate $\Delta_{\text{GT}}\widetilde{S}(t)$. It
is the magnitude of $\Delta_{\text{GT}}\widetilde{S}(t)$ that would determine
whether the entropy of the glass remains above or below $S_{\text{eq}}$ of the
supercooled liquid. \emph{As the entropy change of the medium is completely
reversible, its evaluation will not suffer from any irreversibility going on
in }$\Sigma_{0}$\emph{. }This is the major benefit of investigating a glass as
part of $\Sigma_{0}$.

Both glasses begin to deviate from the equilibrium supercooled liquid at
$T_{0\text{g}}$, but their structures are not yet "frozen;" they freeze over a
long period of time ($t>>\tau_{\text{obs}}$) at a lower temperature
$T_{0\text{G}}$ to form an amorphous solid, to be identified as a glass
(Glass1 and Glass2). The location of $T_{0\text{g}}$\ is determined by the
choice of $\tau_{\text{obs}}$; indeed, $T_{0\text{g}}$ decreases with
increasing $\tau_{\text{obs}}$. Over the transition region between these two
temperatures, the internal variables gradually change form their equilibrium
values at $T_{0\text{g}}$ to their frozen values at $T_{0\text{G}}$.

In the absence of any internal variable, fixing the temperature and pressure
fixes the instantaneous state. As the UCA glass (Glass2) is homogeneous and
has the same constant temperature and pressure as the medium. it appears then
that there cannot be any heat transfer between $\Sigma$ and $\widetilde
{\Sigma}$. If true, then during an isothermal relaxation (constant $T_{0}$ of
the medium under isobaric condition), the first law with $dQ(T_{0}%
,t)\overset{\text{UCA}}{=}0$ and no internal variables yields%
\[
dE(T_{0},t)\overset{\text{UCA}}{=}-P_{0}dV(T_{0},t)\leq0;
\]
This implies that $dV(T_{0},t)\geq0$ during relaxation, which is most
certainly not a rule in glasses.

Another important aspect of glass transition, as has become apparent from
recent work, is the violation of the principle of detailed balance and of the
fluctuation disspation theorem \cite{Ritort}, such as the equivalence of the
heat capacity with enthalpy fluctuations. But the theorem is shown to be valid
in UCA \cite{Mauro}, which is quite surprising.

Faced with these hard to understand consequences and the conflict with our own
results, we decided to examine the basic assumptions in the unconventional
approach. These assumptions, to the best of our reading of their work, are not
properly and adequately justified so far by GMc. In many cases, they are
simply stated as facts alongside several statements that are either
exaggerations or are outright false. Therefore, we will treat them as
conjectures and investigate whether we can justify them either rigorously or
on physical grounds. We defer to the next section these conjectures and the
role they play in the logical development of UCA.

We should mention at this point that some aspects of UCA have already been
criticized by other authors
\cite{Goldstein,Nemilov,GujratiResidualentropy,Guj-Comment1,Guj-Comment2,Gujrati-Symmetry,Johari}%
. In particular, Goldstein \cite{Goldstein}, see also
\cite{Guj-Comment1,Guj-Comment2}, demonstrated that the entropy loss during
the glass transition violates the second law. To this GMc responded by
suggesting that the process of glass formation is not governed by the second
law \cite{Gupta0}. This is a surprising response (as the second law is
supposed to govern all processes), but quite understandable as GMc have a very
unconventional view of the second law. We do not get into this debate by
avoiding the issue altogether. We focus on an isolated system $\Sigma_{0}$,
where there cannot be any dispute about the second law; see Eq.
(\ref{Second_Law}). In that sense, our work differs from other attempts
\cite{Goldstein,Nemilov,Johari}. In our investigation, which is at a
fundamental level, we look at all the underlying assumptions of UCA to see if
they can be justified so that UCA could become an acceptable theory. We only
consider the thermodynamic entropy during this part of our investigation, so
we do not get confused by which statistical entropy formulation is appropriate
to study vitrification. Once, we settle the issues by using the thermodynamic
entropy, we turn to the statistical formulation of entropy to assess the
notion of statistical entropy GMc have advocated. It is our belief that the
unconventional view of the second law and of the statistical entropy form the
basis of UCA, which has been justified in various publications by following
the logical steps listed below:

\begin{enumerate}
\item[UCA1.] The use of equilibrium thermodynamics using calorimetric data
cannot determine the entropy of the glass.

\item[UCA2.] Thus, there is no reason for the residual entropy to exist at
absolute zero.

\item[UCA3.] The entropy of a single microstate is zero. As a glass is in one
microstate at absolute zero, its entropy must be zero in accordance with the
third law.

\item[UCA4.] The glass undergoes spontaneous relaxation during which the
entropy increases and reaches that of the equilibrated supercooled liquid
given by the solid curve in Fig. \ref{Fig_entropyglass} from below.

\item[UCA5.] The entropy drop due to the loss of ergodicity and the
spontaneous relaxation with entropy increase are, therefore, inverse processes.

\item[UCA6.] A calculation method for the entropy is developed to show drop in
the entropy during the glass transition region, see Glass2 in Fig.
\ref{Fig_entropyglass} so that the calculated entropy shows no residual
entropy at absolute zero.
\end{enumerate}

It is important to understand their final conclusion, the so-called inverse
relationship (UCA5) and to see if it, and all of its underlying assumptions
(UCA1-UCA4), are consistent with the second law, the only fundamental law of
Nature that is accepted by all including GMc. Various conjectures leading to
UCA5 seem not to be adequately answered so far by GMc. This deficiency by
itself does not mean that UCA is unfounded, but it does mean that it requires
closer scrutiny, which forms the basis of this investigation. These authors
invariably consider their system $\Sigma$\ (the glass) at fixed $T_{0}$ and
$P_{0}$, which means that it is \emph{not} an isolated system; rather, it is
surrounded by $\widetilde{\Sigma}$; see Fig. \ref{Fig_Systems}. (In the
following, we will call their system $\Sigma$ an open system, knowing very
well that this is not the customary usage. We believe that this will not cause
any confusion.) As GMc constantly appeal to the second law in terms of the
entropy of the system, the most convenient way to examine their approach is to
focus on the isolated system $\Sigma_{0}$ in which the glass will be a
possible state of the system $\Sigma$. This allows us to examine their
approach at the most fundamental level.

\subsection{Summary of Results}

We summarize our conclusions that follow from the application of the second
law in terms of entropy to an isolated system. We only consider vitrification
in the rest of the work. We agree with UCA1, but we find UCA2 unsubstantiated.
Indeed, we find the calorimetrically obtained $S_{\text{expt}}(T_{0})$ forms a
\emph{lower bound} to the entropy $S(T_{0})$ so that the residual entropy has
a lower bound $S_{\text{expt}}(T_{0})$ at absolute zero. The latter entropy is
usually non-negative, so that the residual entropy must be even larger than
this. Such a glass cannot satisfy the third law, which leads to UCA3 being
invalid. Indeed, if there is ever any conflict between the second and the
third law, it is the former that supersedes. In an isothermal relaxation, the
entropy actually decreases towards the equilibrated supercooled liquid
entropy, thus invalidating UCA4. However, the irreversible entropy generation
remains non-negative in accordance with the second law. GMc do not recognize
the importance of the irreversible entropy generation for the second law and
mistakenly ascribe its \emph{universal }non-negative\emph{ }aspect to the
entropy of the system. Following UCA, we find that $\widehat{S}(t)=0$ all
times including $t=0$, when the external condition (such as the temperature)
of the system is changed. This is inconsistent with UCA5. Even if we follow
UCA Conjecture \ref{Marker_Microstate}, see Sect. \ref{Sect_UCA_Conjectures},
although it is inconsistent with the first part, we find that the entropy now
increases at $t=0$ and reaches that of Glass2 at $t=t_{\text{obs}}$; the glass
transition occurs at this instant if the external condition of the system is
disturbed somehow. If it is not disturbed, the entropy would continue to
increase. Thus the entropy is always increasing for $t\geq0$ in UCA. In both
cases, we do not find that there is any justification in calling the glass
transition and relaxation to be inverse processes in UCA. Thus, UCA5 is not a
consequence of the previous steps UCA1-UCA4. Just because GMc have provided a
computational scheme to support their invalid conclusion cannot be considered
a proof of the validity of UCA. We find that UCA misses out many important
aspects of non-equilibrium systems such as their temperature, pressure, etc.
being different from those of the medium, absence of any internal variables to
capture additional irreversible entropy generation, the failure of the
fluctuation-dissipation theorem etc.

\section{Important Conjectures in UCA\label{Sect_UCA_Conjectures}}

As the recent work from GMc is expected to represent their most up-to-date and
current state of understanding of various issues, we will mostly focus on
their recent work \cite{Mauro} for an understanding of the technical aspect of
their approach, the cornerstone of which is that it treats structural
relaxations as the \emph{inverse} of the glass transition \cite{Mauro}:

\begin{conjecture}
\label{Marker_Inverse}The inverse of the glass transition is structural
relaxation, which involves a restoration of ergodicity as a glass
spontaneously approaches the liquid state. This spontaneous relaxation process
is called a unifying process and must entail an increase in entropy as the
observation time constraint is lifted.
\end{conjecture}

The decrease in entropy during glass transition to that of Glass2 is justified
on the basis of a seemingly innocuous conjecture about the effect of
confinement to an ergodic component stated as a \emph{fact }\cite{Mauro}:

\begin{conjecture}
\label{Marker_entropy_loss}The loss of ergodicity at the glass transition
necessarily involves a loss of configurational entropy, since this causes the
system to be confined to a subset of the overall phase space. At absolute
zero, any glass is confined to one and only one microstate, so the
configurational entropy of a glass is necessarily zero, in accordance with the
Third Law and the principle of causality.
\end{conjecture}

In vitrification, the entropy does decrease. This is most clearly seen by
quenching the supercooled liquid from a temperature just above $T_{0\text{g}}$
to A below $T_{0\text{g}}$; see Fig. \ref{Fig_entropyglass}. The decrease is
shown by the dashed downward arrow to B, which represents the entropy of
Glass1 that stays above $S_{\text{eq}}$. A similar drop of much higher value
occurs for Glass2. Both glasses seem to conform to the first part of the
conjecture. Therefore, this part cannot be the defining characteristics of
ergodicity loss or UCA. It is merely a consequence of a positive heat capacity
and nothing more and has nothing to do with ergodicity loss.\ Whether the loss
is big enough to satisfy Eq. (\ref{Entropy_Behavior_UCA}) is never
demonstrated as they have not calculated the entropy gain $\Delta_{\text{GT}%
}\widetilde{S}(T_{0},t)$ of the medium. The second half of the conjecture
requires the entropy to vanish at absolute zero, so the entropy of the UCA
glass is given by the dashed curve Glass2 in Fig. \ref{Fig_entropyglass} and
not by the dotted curve Glass1. Thus, UCA requires the entropy to drop below
$S_{\text{eq}}$ of the equilibrated supercooled liquid and requires evaluating
$\Delta_{\text{GT}}\widetilde{S}(T_{0},t)$ to substantiate it. This part of
the conjecture is based on the following conjecture \cite{Mauro0}:

\begin{conjecture}
\label{Marker_Microstate} \ldots an instantaneous measurement \ldots causes
the system to \textquotedblleft collapse\textquotedblright\ into a single
microstate $i$ with probability $p_{i}(t)$. In the limit of zero observation
time, the system is confined to one and only one microstate and the observed
entropy is necessarily zero. However, the entropy becomes positive for any
finite observation time $\tau_{\text{obs}}$\ since transitions between
microstates are not strictly forbidden except at absolute zero, barring
quantum tunneling.
\end{conjecture}

The conjecture refers to a system "collapsing" into one out of \emph{many}
microstates and asserts as a fact that \emph{the entropy of a microstate is
identically zero} without any supporting justification. This creates some
conceptual problems. At each instant of time $t$, any system, not necessarily
a glass only, is going to be in some microstate $i_{t}$; of course, we do not
know which microstate it would be in at that instance. If we make "an
instantaneous measurement," the system will remain in that microstate; there
is no "collapse" of the microstate. Even the macrostate, which is by
definition the collection of all relevant microstates along with their
probabilities, does not have time to change, because the probability
distribution $p_{i}(t)$ does not change. Indeed, one does not need to make any
measurement on the system to conclude "\ldots\textit{the system to collapse
into a single microstate\ldots}." At each instance, the system is going to be
in some microstate. If we interpret a measurement as something that
instantaneously alters the external condition such as the fields of the
medium, then such an instantaneous move (measurement) will probe the
instantaneous microstate of the system. The statistical entropy of the system
$\widehat{S}(t)=0$ at each instant, and therefore at all times, in accordance
with Conjecture \ref{Marker_Microstate}. This is true even if the system is an
isolated system not in equilibrium. This concept of statistical entropy in UCA
is in direct contradiction with the second law in Eq. (\ref{Second_Law})
according to which the thermodynamic entropy $S_{0}(t)$ is not constant in
time. Thus, the conjecture needs some justification, which GMc have not
provided so far. In particular, it allows us to make the following

\begin{remark}
\label{Marker_Remark_GMc_entropy}\emph{The statistical concept of entropy used
by GMc in the above conjecture has nothing to do with the thermodynamic
entropy used in Eq. (\ref{Second_Law}).}
\end{remark}

Another conceptual problem is that the statistical entropy is defined for a
macrostate as an \emph{average} quantity over all microstates; see Eq.
(\ref{Gibbs_Entropy}). If it happens that a certain macrostate consists of a
single microstate whose probability must be $p(t)=1$, then the statistical
entropy of that macrostate such as a completely ordered crystal is certainly
zero. But what does one mean by the statistical entropy, an average quantity,
of a microstate? What averaging does one perform for a microstate? \ This
issue is never addressed by GMc except by the above conjecture and by an
appeal to the Boltzmann entropy formulation $\widehat{S}(t)=\ln W(t)$ (we set
$k_{\text{B}}=1$), which requires the number of microstates $W(t)$ forming the
macrostate. This is evident from \cite{Mauro}

\begin{conjecture}
\label{Marker_Boltzmann}\ldots Boltzmann's definition of entropy is the only
one valid and consistent with the Second Law for non-equilibrium system.
\end{conjecture}

They take $W(t)=1$ for a single microstate at all temperatures, even if there
may be other possible microstates, and argue for zero entropy. Thus, the
instantaneous statistical entropy will always be zero at all times and at all
temperatures since the system is in a single microstate at each instant. The
idea of introducing an instantaneous measurement is highly appropriate as we
need to measure instantaneous values of the observables. No measuring
apparatus will ever measure the instantaneous entropy; its value can only be
inferred indirectly. Thus, entropy is not an observable in the same sense as
the mechanical variables such as energy, volume, etc. are; it is a
thermodynamic quantity, which has been given a statistical interpretation in
statistical mechanics. Despite this, GMc argue that, when the measurement
takes some non-zero time, the statistical entropy increases with the duration
of measurement. However, GMc never clarify if the measurement gives an
accumulated value or the average value of any quantity. The first option is
counter-intuitive as this suggests that the value of the energy by such a
measurement will increase with the duration of the measurement. The second
option seems reasonable as the value of the measurement will give an average
energy. This then suggests that, since at each instant during the measurement,
the system is in a single microstate so that its entropy is zero, the
measurement will still result in a zero entropy. Why does it increase? Even if
we adopt the first option, then the "measured" entropy would still be zero as
accumulating zero always gives zero. No explanation is offered by GMc for this
part of the conjecture. Recall that one cannot appeal to the second law, which
uses the thermodynamic entropy, while GMc use their statistical entropy whose
equivalence with the thermodynamic entropy is never shown by them. We have
addressed this issue elsewhere
\cite{GujratiResidualentropy,Gujrati-Symmetry,Guj-Recurrence,Guj-Irreversibility}
with a very different conclusion. We find, see Sect.
\ref{Sect_Statistical_Entropy}, that the statistical entropy contribution of a
microstate $i$\ is $-\ln p_{i}(t)$, which is inconsistent with the above
conjecture.\ Thus, we need to understand the basis of their conjecture. This
conjecture is amplified by the following two inter-connected Conjectures
\cite{Gupta}:

\begin{conjecture}
\label{Marker_Boltzmann_Spontaneous}With Boltzmann's definition, the entropy
increases during a spontaneous process.

\begin{conjecture}
\label{Marker_Glass_Nonspontaneous}If entropy increases during the relaxation
process and the glass transition is the inverse of relaxation, then entropy
must decrease during the glass transition. This is consistent with our
previous conclusion that the glass transition is nonspontaneous. While this
conclusion may appear as inconsistent with the second law, there is no
violation since the second law is a statement about spontaneous processes,
i.e., processes in which a system relaxes toward an equilibrium or less
constrained state. The glass transition is not such a process since here an
equilibrium system becomes a constrained equilibrium state.
\end{conjecture}
\end{conjecture}

Conjecture \ref{Marker_Boltzmann_Spontaneous} is not only inconsistent with
Conjecture \ref{Marker_Microstate}, it is inconsistent with classical
thermodynamics when we consider a system which is not isolated. For such a
system, its relevant free energy decreases in any spontaneous process. It
appears that GMc confuse isothermal relaxation occurring in a glass with
spontaneous processes occurring at fixed observables such as energy, volume,
etc. The latter processes occur for isolated systems, not for a glass at fixed
temperature and pressure. If GMc insist on focusing on the entropy, then the
statement should be in terms of the irreversible entropy generation
$\Delta_{\text{i}}S(t)\geq0$, not in terms of the entropy change $\Delta
S(t)$. Thus, for the conjecture to make sense, the reversible entropy drop
$\Delta_{\text{e}}S(t)$ during vitrification must not be too negative to
ensure $\Delta S(t)>0$. This requires a justification that%
\[
\left\vert \Delta_{\text{e}}S(t)\right\vert <\Delta_{\text{i}}S(t)
\]
for the conjecture to be valid. However, no such justification is offered by
GMc in their work. As given, it gives the impression that UCA treats entropy
to increase during relaxation in all kinds of systems, isolated or not. This
is unsettling. The same problem occurs with the last conjecture.

Therefore, to determine whether the above conjectures are justifiable, we turn
to the second law for $\Sigma_{0}$.

\section{Consequences of the Second Law for $\Sigma_{0}$%
\label{Sect_Second_Law}}

\subsection{Irreversible Entropy Generation}

The second law does tell us that the irreversible entropy generation in any
spontaneous process is non-negative, but leaves the behavior of the entropy
undetermined; the latter depends on the process. The entropy that appears in
the second law in classical thermodynamics is a thermodynamic concept.
\emph{It is postulated to exist even when the system is not in equilibrium;
its existence and continuity neither requires any statistical interpretation
nor does it require the third law. If there is any conflict between the second
law and any other laws of physics, the second law will always win. }In
general, in any thermodynamic process from macrostate 1 to macrostate 2, the
change in the entropy\cite{Donder,deGroot,Prigogine,Guj-NE-I,Guj-NE-II}%
\begin{equation}
\Delta S\equiv S_{\text{2}}-S_{\text{1}}=\Delta_{\text{e}}S+\Delta_{\text{i}%
}S,\ \ \Delta_{\text{i}}S\geq0,\label{Entropy_Change}%
\end{equation}
in which $\Delta_{\text{i}}S$ denotes the irreversible entropy generation
within the system and $\Delta_{\text{e}}S$ denotes the reversible entropy
change due to exchange with the medium. The actual value of $\Delta S$ will
depend on the values of $\Delta_{\text{e}}S$ and $\Delta_{\text{i}}S,$ and can
have any sign. In general, we have
\begin{equation}
\Delta S\geq\Delta_{\text{e}}S,\label{Entropy_inequality}%
\end{equation}
which will prove extremely important below. In the following, we will only be
interested in vitrification for which  $\Delta_{\text{e}}S$ is negative. If it
happens that $\Delta_{\text{e}}S$ is negative enough to overcome the positive
contribution of $\Delta_{\text{i}}S$, then we will obtain a negative $\Delta
S$.

\subsection{Second Law for $\Sigma_{0}$}

For an isolated system such as $\Sigma_{0}$, $\Delta_{\text{e}}S=0$ so it is
not surprising that%
\begin{equation}
\Delta S\overset{\text{isolated system}}{=}\Delta_{\text{i}}S\geq0,
\label{Second_Law_Isolated}%
\end{equation}
which explains the standard formulation \cite{note1} of the second law but
only for an isolated system. We now consider $\Sigma_{0}$, but we allow the
system to be not in equilibrium with the medium. The medium is at a fixed
temperature $T_{0}$ and pressure $P_{0}$. We are thinking of the system that
has been brought in contact with the medium at some instant $t=0$, which we
then follow in time. The entropy of $\Sigma_{0}$ is written as a sum of the
entropies of the system $S(t)$ and the medium $\widetilde{S}(t)$:%
\begin{equation}
S_{0}(t)=S(t)+\widetilde{S}(t). \label{Entropy_Sum}%
\end{equation}
During its approach towards the maximum, the instantaneous temperature,
pressure, etc. of the system, if they can be defined, are different from those
of the medium \cite{Guj-NE-I,Guj-NE-II,Langer}. The condition required for
defining temperature, pressure, etc. of any non-equilibrium system is that its
entropy is a function of its instantaneous observables and internal variables
\cite{Donder,deGroot,Prigogine,Guj-NE-I,Guj-NE-II} only; it has no explicit
time-dependence. The system is said to be in \emph{internal equilibrium
}\cite{Guj-NE-I,Guj-NE-II}, when this condition is met. Unless this condition
is met, we cannot identify fields for the system, even though they exist for
the medium. For simplicity, we will consider a system with a fixed number of
particles with only observables $E(t)$ and $V(t)$ along with just one internal
variable $\xi(t)$. Thus,
\begin{equation}
S(t)\equiv S(E(t),V(t),\xi(t)). \label{Entropy_Function}%
\end{equation}
At each instance, $E(t),V(t)$ and $\xi(t)$ depend on the history of the
system. The corresponding fields are now given by respective derivatives of
the entropy:%
\[
\frac{1}{T(t)}=\left(  \frac{\partial S(t)}{\partial E(t)}\right)
,~\ \ \frac{P(t)}{T(t)}=\left(  \frac{\partial S(t)}{\partial V(t)}\right)
,\ \frac{A(t)}{T(t)}\equiv\left(  \frac{\partial S(t)}{\partial\xi(t)}\right)
;
\]
the new variable $A(t)$ represents the affinity conjugate to $\xi(t).$ The
Gibbs fundamental relation is given by%
\begin{equation}
dS(t)=\frac{1}{T(t)}dE(t)+\frac{P(t)}{T(t)}dV+\frac{A(t)}{T(t)}d\xi(t),
\label{Gibbs_Fundamental(t)0}%
\end{equation}
which can be rewritten as
\begin{equation}
dE(t)=T(t)dS(t)-P(t)dV(t)-A(t)d\xi(t) \label{First_Law(t)0}%
\end{equation}
for the non-equilibrium system. In equilibrium, this relation will reduce to%
\begin{equation}
dE=T_{0}dS-P_{0}dV, \label{First_Law_Eq}%
\end{equation}
where all variables are independent of time and we have used the fact that the
equilibrium value $A_{0}$ of $A(t)$\ vanishes. We can rewrite Eqs.
(\ref{Gibbs_Fundamental(t)0}) and (\ref{First_Law(t)0}) in the following form
\cite{Guj-NE-II}%
\begin{align}
dS(t)  &  =\frac{1}{T_{0}}dE(t)+\frac{P_{0}}{T_{0}}dV+\left[  \frac{1}%
{T(t)}-\frac{1}{T_{0}}\right]  dE(t)+\left[  \frac{P(t)}{T(t)}-\frac{P_{0}%
}{T_{0}}\right]  dV+\frac{A(t)}{T(t)}d\xi(t),\label{Gibbs_Fundamental(t)}\\
dE(t)  &  =T_{0}dS(t)-P_{0}dV(t)+[T(t)-T_{0}]dS(t)-[P(t)-P_{0}]dV(t)-A(t)d\xi
(t). \label{First_Law(t)}%
\end{align}
The last three terms in Eq. (\ref{Gibbs_Fundamental(t)}) each give three
distinct irreversible entropy generation terms, and must be individually
\emph{non-negative} in accordance with the second law. Let us consider the
middle term in this equation, which is non-negative. In a vitrification
process, the energy of the system decreases so that $dE(t)$ is negative$.$
Thus, in a vitrification process,
\begin{equation}
T(t)>T_{0} \label{Temperature_Relation}%
\end{equation}
during isothermal relaxation (constant $T_{0}$) and approaches $T_{0}$ from
above as the relaxation ceases after equilibrium is achieved \cite{Guj-NE-I}.
As this is a general result coming from the second law, it must be valid for
all non-equilibrium systems including glasses. We need to see whether both
glasses shown in Fig. \ref{Fig_entropyglass} satisfy this result.

For the enthalpy $H(t)\equiv E(t)+P_{0}V(t)$, we find%
\begin{equation}
dH(t)=T_{0}dS(t)+V(t)dP_{0}+[T(t)-T_{0}]dS(t)-[P(t)-P_{0}]dV(t)-A(t)d\xi(t).
\label{Enthalpy_Differential}%
\end{equation}

Let us consider the consequences of the second law. From now on, we focus on
isobaric processes carried out at a fixed pressure $P_{0}$ of the medium. We
will assume that in such an isobaric process, $P(t)=P_{0}$ at all times. For
the time derivative of the entropy $S_{0}$ of the isolated system at fixed
$T_{0}$, it can be shown \cite{Guj-NE-I,Guj-NE-II} that in terms of the
enthalpy and the internal variable of the system%
\begin{equation}
\frac{dS_{0}(t)}{dt}=\left(  \frac{1}{T(t)}-\frac{1}{T_{0}}\right)
\frac{dH(t)}{dt}+\frac{A(t)}{T(t)}\frac{d\xi(t)}{dt}\geq0.
\label{Total_Entropy_Rate}%
\end{equation}
Each term on the right side of the first equation gives an irreversible
entropy generation, see Eq. (\ref{Second_Law_Isolated})), and must be
non-negative. Accordingly,
\begin{equation}
\left(  \frac{1}{T(t)}-\frac{1}{T_{0}}\right)  \frac{dH(t)}{dt}\geq0.
\label{Total_Entropy_Rate1}%
\end{equation}
which is unaffected by the number\ of internal variables. With Eq.
(\ref{Temperature_Relation}), this shows that
\begin{equation}
dH(t)/dt\leq0, \label{Enthalpy_rate}%
\end{equation}
which is found to hold in vitrification.

\subsection{Determination of $\Delta_{\text{GT}}S(T_{0},t_{\text{obs}})$ and
$\Delta_{\text{GT}}\widetilde{S}(T_{0},t_{\text{obs}})$}

From Eqs. (\ref{Gibbs_Fundamental(t)}) and (\ref{Total_Entropy_Rate}), we find
that%
\begin{equation}
\frac{dS(t)}{dt}=\frac{1}{T(t)}\frac{dH(t)}{dt}+\frac{A(t)}{T(t)}\frac
{d\xi(t)}{dt},\ \ \frac{d\widetilde{S}(t)}{dt}=-\frac{1}{T_{0}}\frac
{dH(t)}{dt}; \label{Entropy_variation1}%
\end{equation}
the two equations give the rate at which the entropy of the system and of the
medium change. In vitrification, the rate for the medium is positive. The
second term in the entropy rate for the system is non-negative.

The rate of the entropy drop is given by%
\[
\left\vert \frac{dS(t)}{dt}\right\vert =\frac{1}{T(t)}\left\vert \frac
{dH(t)}{dt}\right\vert -\frac{A(t)}{T(t)}\frac{d\xi(t)}{dt}\leq\frac{1}%
{T(t)}\left\vert \frac{dH(t)}{dt}\right\vert \leq\frac{d\widetilde{S}(t)}%
{dt},
\]
so that the rate of entropy drop for the system is bounded from above by the
rate of entropy gain of the medium. Thus, the drop $\left\vert \Delta
S\right\vert $ for $\Sigma$ is bounded from above by the entropy gain
$\Delta\widetilde{S}$ during some interval $\Delta t$:%
\[
\left\vert \Delta S\right\vert \leq\Delta\widetilde{S},
\]
where we have introduced the change in any thermodynamic quantity $\digamma$:%
\[
\Delta\digamma(t)\equiv\digamma(t)-\digamma(t=0).
\]
From Eq. (\ref{Entropy_variation1}), we find that%
\[
\Delta_{\text{e}}S(t)=\frac{1}{T_{0}}\Delta H(t),\ \ \Delta_{\text{i}%
}S(t)=\int_{0}^{t}\frac{\overset{\cdot}{H}(t)}{T(t)}dt-\frac{\Delta
H(t)}{T_{0}}+\int_{0}^{t}\frac{A(t)\overset{\cdot}{\xi}(t)dt}{T(t)}%
,\ \ \Delta_{\text{e}}\widetilde{S}(t)=-\frac{1}{T_{0}}\Delta H(t),
\]
where the dot above a symbol represents the time-derivative. The maximum
entropy drop $\left\vert \Delta S\right\vert $ occurs when $\Sigma\ $comes to
equilibrium with the medium. In this case, $T(t)\rightarrow T_{0}$ and
$A(t)\rightarrow0$.\ It is also equal to the maximum entropy gain of the
medium at equilibration. This situation correspond to the system as the
equilibrated supercooled liquid with its entropy $S_{\text{eq}}$ given by the
solid curve in Fig. \ref{Fig_entropyglass}. Thus, we conclude that the entropy
of the non-equilibrium system always stays above that of the supercooled
liquid%
\[
S(T_{0},t)\geq S_{\text{eq}}(T_{0})
\]
during vitrification. During relaxation, $S(T_{0},t)\ $approaches
$S_{\text{eq}}(T_{0})$ from above so that
\begin{equation}
dS(t)/dt\leq0; \label{Entropy_variation0}%
\end{equation}
this conclusion is valid regardless of the number of internal variables. The
equality occurs only when equilibrium with the medium has been achieved.

We can understand this result on physical grounds as follows. Let
$T_{0}^{\prime}>T_{0\text{g}}$ denote the temperature from which the system is
cooled by bringing the system in contact with the medium at $T_{0}%
<T_{0\text{g}}$ at $t=0$; see point A in Fig. \ref{Fig_entropyglass}. This
situation corresponds to a rapid quench. Right after the contact, the system
has not had any time to change its microstate and remains in the microstate it
had just prior to the contact is established. Thus, at $t=0,$ the system has
the older entropy $S_{\text{eq}}(T_{0}^{\prime})$ of the equilibrated
supercooled liquid at the previous temperature. We will justify this fact
later; see Eq. (\ref{Microstate_Entropy}). The equilibrium entropy
$S_{\text{eq}}(T_{0})$ must be lower than $S_{\text{eq}}(T_{0}^{\prime})$
since $T_{0}<T_{0}^{\prime}$. Therefore, the entropy must continue to drop
during relaxation at $T_{0}$ even if the contribution from $\overset{\cdot
}{\xi}(t)$\ is non-negative. This gives us the entropy of Glass1 at
$t=t_{\text{exp}}$. The derivation of the last equation given in
\cite{Guj-NE-I} was simpler as we had not considered the internal variable
$\xi(t)$ there. The above discussion justifies the behavior in the
conventional approach described by Eq. (\ref{Entropy_Behavior_CA}).

From Eqs. (\ref{Enthalpy_rate}), (\ref{Entropy_variation1}) and
(\ref{Entropy_variation0}), we find%
\begin{equation}
\frac{dS(t)}{dH(t)}=\frac{1}{T(t)}\left[  1+A(t)\frac{d\xi(t)}{dH(t)}\right]
\geq0,\label{dS/dH_relation}%
\end{equation}
which proves Eq. (\ref{Temperature(t)}). This equation differs from the one
given in \cite{Guj-NE-I}, see Eq. (\ref{dS/dH_relation_0}) because of the
contribution from the internal variable, which was not considered there.
However, the sign of the ratio remains the same.

From Eq. (\ref{Entropy_variation1}), we find%
\[
\Delta\widetilde{S}(t)\equiv\widetilde{S}(t)-\widetilde{S}(t=0)=-\frac{\Delta
H(t)}{T_{0}}=-\frac{H(t)-H(t=0)}{T_{0}}%
\]
At the glass transition, we then have%
\[
\Delta_{\text{GT}}\widetilde{S}(T_{0},t_{\text{obs}})\equiv-\frac
{H(T_{0},t_{\text{obs}})-H(T_{0},t=0)}{T_{0}}\leq\frac{H(T_{0}%
,t=0)-H_{\text{eq}}(T_{0})}{T_{0}}\equiv\Delta_{\text{eq}}\widetilde{S}%
(T_{0}),
\]
where $\Delta_{\text{eq}}\widetilde{S}(T_{0})$\ is the maximum entropy gain of
the medium, which occurs when the system has come to equilibrium with the
medium. This gain can be easily obtained by using a reversible path from the
initial temperature $T_{0}^{\prime}$ to the final temperature $T_{0}$. This
means that in Fig. \ref{Fig_entropyglass}, the system continues along the
equilibrated supercooled liquid. Thus,%
\[
\Delta_{\text{eq}}\widetilde{S}(T_{0})=-\Delta_{\text{eq}}S(T_{0})\equiv
S_{\text{eq}}(T_{0}^{\prime})-S_{\text{eq}}(T_{0}).
\]
In such a reversible process, $\Delta S_{0}=0$, as expected.

Let us compare this entropy drop with that in the system. Because of the
non-negative contribution from $\overset{\cdot}{\xi}(t)$, the entropy change
$\Delta_{\text{GT}}S(T_{0},t)$ satisfies%
\begin{equation}
\left\vert \Delta_{\text{GT}}S(T_{0},t_{\text{obs}})\right\vert \leq\int
_{0}^{t_{\text{obs}}}\frac{\overset{\cdot}{H}(t)}{T(t)}dt\leq\frac{\left\vert
\Delta_{\text{GT}}H(t_{\text{obs}})\right\vert }{T_{0}}\equiv\Delta
_{\text{GT}}\widetilde{S}(T_{0},t_{\text{obs}})\leq\Delta_{\text{eq}%
}\widetilde{S}(T_{0}). \label{Entropy_Bound_Medium}%
\end{equation}
Thus, the entropy change of the medium and of the system satisfy Eq.
(\ref{GT_Entropies}) at the glass transition. But most importantly, the
entropy loss of the medium cannot exceed $\left\vert \Delta_{\text{eq}}%
S(T_{0})\right\vert $, so that
\begin{equation}
\left\vert \Delta_{\text{GT}}S(T_{0},t_{\text{obs}})\right\vert \leq
\Delta_{\text{eq}}\widetilde{S}(T_{0})=\left\vert \Delta_{\text{eq}}%
S(T_{0})\right\vert . \label{Entropy_Bound_System_Medium}%
\end{equation}
We thus conclude \cite{Guj-NE-I,Guj-NE-II} that the entropy of the glass must
stay above that of the equilibrated supercooled liquid, which makes Glass1 as
the physical glass, a conclusion based on the second law.

We can now extend the discussion to\ continuous cooling as follows. We take
$T_{0}^{\prime}>T_{0\text{g}}$ and $T_{0}^{(1)}=T_{0\text{g}}-\Delta T_{0},$
and wait for $\Delta t=t_{\text{obs}}$. The entropy is that of Glass1 at
$T_{0}^{(1)}$. We now decrease the temperature by $\Delta T_{0}$ to
$T_{0}^{(2)}=T_{0\text{g}}-2\Delta T_{0},$ and wait for $\Delta
t=t_{\text{obs}}$. The entropy is that of Glass1 at $T_{0}^{(2)}$. We follow
this cooling until the entire Glass1 curve is obtained. At each temperature,
the entropy of the glass must stay above $S_{\text{eq}}$ of the equilibrated
supercooled liquid. We have thus proved the following important theorem for
any non-equilibrium system:

\begin{theorem}
\label{Theorem_Relaxation}The entropy of any non-equilibrium system such as a
glass in isobaric cooling must stay above that of the equilibrated state.
\end{theorem}

\subsection{Thermodynamic Entropy and Glasses}

In the above discussion, which starts with the second law behavior of the
thermodynamic entropy $S_{0}(t)$ of $\Sigma_{0}$,\ no assumption about the
form of the thermodynamic entropy $S(t)$ (such as whether $S(t)$ lies above
(Glass1) or below (Glass2) the entropy of the equilibrated supercooled liquid;
see Fig. \ref{Fig_entropyglass}) of or the nature of irreversibility such as
loss of ergodicity, chemical reaction, chaos, phase transition, etc. in the
system is made. We do not impose any statistical interpretation on these
entropies either; they are assumed to exist as thermodynamic quantities in
classical thermodynamics. Thus, their values are not relevant; all that is
relevant is their rate. Accordingly, we do not have to even worry if the
entropies needed to be treated as statistical quantities with certain
particular formulation of entropy (Boltzmann versus Gibbs, or the modification
in UCA). Any attempt to identify these classical entropies statistically must
still conform to the consequence of the second law expressed, for example, in
Eq. (\ref{Entropy_variation0}) during vitrification.

Although no assumption was made regarding $S(T_{0},t)$ lying above or below
the entropy $S_{\text{eq}}(T_{0})$ of the equilibrated supercooled liquid, the
second law has resulted in Glass1 as being the physically relevant glass, and
not Glass2. The deviation of Glass1 entropy $S(T_{0},t)$ from $S_{\text{eq}%
}(T_{0})$ of the solid curve is due to the irreversible contributions. The
entropy of the non-equilibrium state Glass1 approaches that of the
equilibrated supercooled liquid entropy from above during isothermal
relaxation. This downward approach of the entropy of Glass1 is a consequence
of the second law. The entropy of the system, howsoever defined, must satisfy
Eq. (\ref{Entropy_variation0}) in vitrification if the system has to obey the
second law.

We assume that at time $t=0$, the system is above $T_{0\text{g}}$, so that the
system is an equilibrated supercooled liquid. Its temperature and pressure are
equal to those of the medium. Let the $E^{\prime},V^{\prime}$ and $S^{\prime}$
denote the energy, volume, and entropy of the equilibrated supercooled liquid
at this temperature $T_{0}^{\prime}$, respectively. The equilibrium value of
the internal variable is denoted by $\xi^{\prime}$. At time $t=0$, we abruptly
bring this system in \emph{contact} with another medium at temperature $T_{0}$
just below $T_{0\text{g}}$; see for example point A in Fig.
\ref{Fig_entropyglass}. Immediately after the contact, the initial state of
the system is characterized by its observables $E^{\prime},V^{\prime}%
,\xi^{\prime}$ and $S^{\prime}$ at $T_{0}$. After some time $t=t_{\text{exp}}%
$, the system appears to be glassy as shown by B on Glass1. During further
relaxation, the system eventually approaches the equilibrated supercooled
liquid at $T_{0}$. During the relaxation process, the entropy $S(T_{0},t)$
decreases in accordance with Eq. (\ref{Entropy_variation0}). This is an
example of a fast quench. In a continuous vitrification process carried out at
a fixed rate, the resulting glass entropy is shown by Glass1. If such a glass
is allowed to relax, see the two downward arrows, it also converges to the
solid curve of the supercooled liquid. The resulting entropy during heating is
shown by the dash-dotted curve (a) and shows the resulting hysteresis over the
transition region.

With the above background about the role of the second law for the open
system, we turn to UCA to see if we can justify its consequences or assumptions/conjectures.

\section{Irreversible Contributions and Calorimetric Calculation during the
Glass Transition\label{Sect_Glass_Transition}}

GMc assert in the abstract:\cite{Mauro} "\textit{A common assumption in the
glass community is that the entropy of a glass can be calculated by
integration of measured heat capacity curve through the glass transition. Such
integration assumes that glass is an equilibrium material and that the glass
transition is a reversible process...}" This is an inaccurate and highly
misleading statement, which completely overlooks the tremendous progress made
by Prigogine and Defay \cite{Prigogine-Defay}, Davies and Jones \cite{Davies}
and others who have followed the concept of internal variables due to de
Donder \cite{Donder,deGroot,Prigogine,Guj-NE-I,Guj-NE-II}. We refer the reader
to a very nice review by Nemilov \cite{Nemilov} and his monogram
\cite{Nemilov-Book}. Workers in the glass community are well aware of the
fact, see Fig. \ref{Fig_entropyglass}, that the glass transition neither
occurs at a single temperature (it actually occurs over a range $T_{0\text{G}%
}-T_{0\text{g}}$) nor is it reversible (see the dotted curve and the
dash-dotted curve (a) for the entropy during cooling and heating for Glass1).
Any attempt to use equation 1 of GMc \cite{Mauro}, which we slightly modify to
express it in terms of the thermodynamics entropy and present below%
\begin{equation}
\Delta S\equiv S(T_{0})-S(T_{\text{M}})\overset{\text{GMc}}{=}\Delta
_{\text{e}}S\equiv%
{\textstyle\int\limits_{T_{\text{m}}}^{T_{0}}}
\frac{C_{P}(T_{0}^{\prime})}{T_{0}^{\prime}}dT_{0}^{\prime}%
,\label{GM_Entropy_Diff}%
\end{equation}
by workers in the field merely reflects the desire to use an \emph{approximate
}description by replacing $S(T_{0})$ by its calorimetric value $S_{\text{expt}%
}(T_{0})$%
\[
S_{\text{expt}}(T_{0})\equiv S(T_{\text{M}})+%
{\textstyle\int\limits_{T_{\text{m}}}^{T_{0}}}
\frac{C_{P}(T_{0}^{\prime})}{T_{0}^{\prime}}dT_{0}^{\prime}%
\]
the right hand side in the above equation \cite{Mauro}. Comparing with Eq.
(\ref{Entropy_inequality}) shows that the approximation is simply to replace
the forward inequality by a\emph{ forward} \emph{approximate} equality
(approximate equality from the greater side)%
\begin{equation}
\Delta S\gtrapprox\Delta_{\text{e}}S,\label{Entropy_Approximation}%
\end{equation}
and the question one should ask is: \emph{How reliable is the forward
approximation }\cite{Jackel,Nemilov,GujratiResidualentropy,Goldstein}? This
question has also been recently answered by Johari and Khouri \cite{Johari}.
This forward approximation cannot be confused with the above-mentioned "common
assumption" of equality $\Delta S\overset{\text{GMc}}{\equiv}\Delta_{\text{e}%
}S$ in Eq. (\ref{GM_Entropy_Diff}). It appears that GMc confuse this forward
approximation with an equality and use it (see below) to suggest that the
traditional view of glasses is inapplicable \cite{Mauro}. This suggestion is
not the right conclusion.

An important aspect of non-equilibrium systems is that their fields such as
the temperature $T(t)$ are different from the constant fields such as the
temperature $T_{0}$\ of the surrounding medium. Such a two-field description
captures the essence of non-equilibrium states and is also consistent with the
violation of the fluctuation dissipation theorem \cite{Ritort} in
non-equilibrium systems. It has become apparent that non-equilibrium systems
violate the principle of detailed balance and the fluctuation dissipation
theorem. The way irreversibility and the second law \cite{note1} are taken
into account is by instantaneous fields and the introduction of internal
variables; the latter is a standard practice in non-equilibrium thermodynamics
\cite{Nemilov-Book,Donder,deGroot,Prigogine,Guj-NE-I,Guj-NE-II}, but their
role is not considered in UCA. This leads them to make several inaccurate
statements \cite{Mauro}. Not realizing that their Eq. (1) is a forward
inequality \cite{Betsul,Sethna-Paper,Guj-Rigorous}
\[
S(T_{0})\geq S_{\text{expt}}(T_{0})
\]
due to \emph{all} irreversible contributions to the entropy, see the last
three terms in Eq. (\ref{Gibbs_Fundamental(t)}), they incorrectly conclude
"\textit{\ldots\ that glass is treated strictly in the framework of
equilibrium thermodynamics,\ldots}" If the correction is made and the equality
is replaced by a forward inequality, it immediately rules out any
contradiction with the statement "\ldots\textit{that glass is exempt from the
Third Law due to its nonequilibrium nature.\ldots}" since the entropy of the
glass at absolute zero is bounded below by $S_{\text{expt}}(0)$%
\[
S_{\text{res}}\equiv S(0)\geq S_{\text{expt}}(0).
\]
One then discovers that there is no contradiction in logic in the traditional
view, and the following statement \cite{Mauro} in UCA is without any
scientific merit:

\begin{quote}
"\textit{Previous reports of a finite residual entropy of glass at absolute
zero are an artifact of treating glass within the context of equilibrium
thermodynamics or equilibrium statistical mechanics, assuming ergodicity and
without accounting for the observation time constraint.}"
\end{quote}

Not appreciating the important role played by instantaneous fields in
non-equilibrium systems leads them to doubt the applicability of%
\[
dS\equiv dQ(t)/T(t)
\]
due to heat flow to such non-equilibrium systems, where $dQ$ is the heat
transfer with the medium. That it is the correct result follows immediately
from Eq. (\ref{First_Law(t)0}) by rewriting it in the form of the first law
as
\[
dE(t)=dQ(t)-P(t)dV(t)-A(t)d\xi(t)
\]
so that
\[
dQ(t)=T(t)dS(t).
\]
The irreversible entropy generation within the system is given by
$d_{\text{i}}S\equiv dQ\left\{  1/T(t)-1/T_{0}\right\}  \geq0$. In
vitrification, $dQ<0$, which then requires $T(t)\geq T_{0}$. This yields
$dS\geq d_{\text{e}}S\equiv dQ/T_{0}$, which results in $S(T_{0})\geq
S_{\text{expt}}(T_{0})$ as noted above. The equality occurs only in
equilibrium. However, GMc \cite{Mauro} confuse the forward approximate
equation 1 of GMc \cite{Mauro}, reproduced here as Eq. (\ref{GM_Entropy_Diff}%
), with an equality and mistakenly conclude that the classical view is
inapplicable. The conclusion is without any foundation. The second law clearly
establishes that residual entropy is real. As a non-zero residual entropy is
in conflict with the third law can only mean that the third law is not
applicable to non-equilibrium systems, a conclusion well known in theoretical
physics \cite{Landau}.

As GMc do not consider any internal variable, we must consider Eq.
(\ref{dS/dH_relation}) by setting $A(t)=0$. In that case, we have
\begin{equation}
\left(  \frac{\partial S(t)}{\partial H(t)}\right)  _{P}=\frac{1}%
{T(t)}\overset{\text{UCA}}{=}\frac{1}{T_{0}}>0,\label{dS/dH_relation_0}%
\end{equation}
since the temperature of the system is taken equal to $T_{0}$ in UCA. We then
conclude that the increase in their statistical entropy along with the
decrease in enthalpy during relaxation violates the positivity of the
instantaneous temperature of the system and throws doubts that their
statistical entropy can be identified with the thermodynamic entropy. As the
nature of the statistical entropy is crucial to understand the reasons for the
possible failure of $\widehat{S}(t)$, we turn to this issue in the following section.

\section{Statistical Entropy for Non-equilibrium
Systems\label{Sect_Statistical_Entropy}}

\subsection{Statistical Entropy as an average for a Macrostate}

The discussion so far has been about the thermodynamic entropy and its
existence as used in classical thermodynamics and in the formulation of the
second law for an isolated system; see Eqs. (\ref{Second_Law_Isolated}) and
(\ref{Second_Law}). All that is required for this is the fact that \emph{there
exists an entropy function }$S_{0}(t)$\emph{, which is non-decreasing in
time}. Its actual value is not relevant; all that is relevant is the change in
this function. In other words, the thermodynamic entropy is not constrained by
the third law in any way; the latter becomes relevant only for the statistical
interpretation of the thermodynamic entropy. The latter does not even have to
be non-negative, as is well known from the entropy of an ideal gas at low
temperatures. The existence of $S_{0}(t)$ immediately leads to the existence
of $S(t)$ and $\widetilde{S}(t)$; see Eq. (\ref{Entropy_Sum}). However, it
should be emphasized that whatever value any of these entropies such as $S(t)$
has at any instance, it has this value even if no measurement is made on the system:

\begin{remark}
\emph{Any statistical interpretation of the entropy must obey the property
that its value must be unaffected by the measurement. }
\end{remark}

This point should not be overlooked. We will explain later how this statement
is justified in classical thermodynamics or non-equilibrium statistical
dynamics. 

We now turn to the statistical interpretation of entropy that provides a
justification of the third law for equilibrium states and endow the entropy
such as $S_{0}(t)$ with a definite value. Let $i$ denote a microstate of the
isolated system $\Sigma_{0}$ in some macrostate. The formulation by Gibbs in
terms of the probability $p_{i}(t)$ of a microstate $i$ at time $t$ is as
follows:%
\begin{equation}
S_{0}(t)\equiv-%
{\textstyle\sum\limits_{i}}
p_{i}(t)\ln p_{i}(t)\equiv-\left\langle \eta(t)\,\right\rangle
,\label{Gibbs_Entropy}%
\end{equation}
where the sum is over all microstates, whose number is $W_{0}$, associated
with the particular macrostate; we have set $k_{\text{B}}=1$. A microstate is
called \emph{available} at time $t$ if its probability is non-zero; otherwise,
it is \emph{unavailable} at that time \cite{Gujrati-Symmetry}. An available
microstate does not mean that the microstate has necessarily been visited by
the system during the time interval $t$. Following Gibbs \cite{Gibbs}, we have
introduced the \emph{index of probability}%
\[
\eta(t)\equiv\ln p(t),
\]
so that the entropy becomes a statistical average of the negative index of
probability \emph{over} all microstates belonging to the macrostate. This
makes entropy similar to any other average mechanical observable like the
energy:%
\begin{equation}
E_{0}(t)\equiv%
{\textstyle\sum\limits_{i}}
p_{i}(t)E_{0i},\label{Average_Energy}%
\end{equation}
where $E_{0i}$ is the energy of the $i$-th microstate. The only difference is
that the entropy is a thermodynamic quantity as an average of $-\eta(t)$. The
index has its origin in the stochastic nature
\cite{Guj-Recurrence,Guj-Irreversibility,Gujrati-Symmetry} of a statistical
system. Thus, its nature is very different from the mechanical nature of
observables like the energy, momentum, etc. although both averages give a
statistical average. It is clear from Eq. (\ref{Gibbs_Entropy}) that the
negative index $-\eta(t)$ is the contribution to the entropy from a single
microstate. One may wish to think of $-\eta(t)$ as the entropy of a
microstate, but this is not the conventional view as the entropy is an average
quantity for the macrostate; see, however, Eq. (\ref{Microstate_Entropy}). We
refer the reader to recent reviews for more details
\cite{GujratiResidualentropy,Gujrati-Symmetry}.

The probabilities $p_{i}(t)$ can be determined by considering an ensemble or
by considering the temporal evolution, as described at length elsewhere
\cite{Guj-Irreversibility,Guj-Recurrence,GujratiResidualentropy,Gujrati-Symmetry}%
, but neither is really necessary provided the probabilities $p_{i}(t)$ are
known. If it is known initially that $\Sigma_{0}$ is in some \emph{unique}
microstate $i_{0}$, then $p_{i}(0)\equiv\delta_{i,i_{0}}$ and $S_{0}(0)=0.$ As
time goes on, and assuming that the dynamics is \emph{stochastic}, the initial
state will result in making various microstates available with some
probabilities $p_{i}(t)$ at time $t$, and the entropy given by Eq.
(\ref{Gibbs_Entropy}) will increase \cite{Tolman}, until it reaches its
maximum value. It is most certainly not a constant
\cite{Guj-Irreversibility,Guj-Recurrence,Gujrati-Symmetry}.

The time needed for all the microstates to be available is, in most cases,
much shorter than the Poincar\'{e} recurrence
time\ \cite{Guj-Irreversibility,Guj-Recurrence,Gujrati-Symmetry,GujratiResidualentropy}%
. It may indeed be smaller than the relaxation time $\tau_{\text{relax}}$. At
the shorter time, all microstates have become available, but the entropy is
still not necessarily at its maximum for the macrostate, since the microstates
are not yet \emph{equiprobable}. In the latter situation, the system is in
internal equilibrium to be discussed below. If and only if \emph{all}
microstates are equally probable ($p_{i}(t)\rightarrow1/W_{0}$ for all $i$),
which happens after the relaxation time $\tau_{\text{relax}}$, do we have the
maximum possible value of the entropy for the equilibrium macrostate:%
\begin{equation}
S_{0}(t)\rightarrow S_{0\text{,eq}}\equiv\ln W_{0}\ \ \ \ \text{for }%
t\gtrsim\tau_{\text{relax}}.\label{Boltzmann_Entropy}%
\end{equation}
This entropy is known as the Boltzmann entropy. It is the equilibrium value of
the entropy and occurs because \emph{all} microstates of the system are
equiprobable. This entropy is constant in time and depends on the constant
observables $E_{0},V_{0},N_{0}$,~etc. The internal variables that are now
constant are not independent of these observables in equilibrium
\cite{Guj-NE-I,Guj-NE-II}, so the entropy does not depend on them anymore.

It is possible in many cases that over a period of time smaller than
$\tau_{\text{relax}}$, only a part of microstates, whose number is given by
$W_{0}(t)<W_{0}$ have become available. In that case, the sum in Eq.
(\ref{Gibbs_Entropy}) is restricted to $W_{0}(t)$. But the entropy is strictly
less than $\ln W_{0}(t)$ unless all available microstates become equally
probable. In that case, the entropy is given by
\begin{equation}
S_{0}(t)\rightarrow S_{0\text{in.eq}}\equiv\ln W_{0}(t)\ \ \ \text{\ for
}t<\tau_{\text{relax}},\label{Boltzmann_Entropy(t)}%
\end{equation}
which is the Boltzmann entropy at that instance. This entropy is now a
function of the internal variables, which themselves depend on time. However,
$S_{0\text{in.eq}}(E_{0},V_{0},N_{0},\xi_{0}(t))$ cannot have an explicit
time-dependence as for fixed $E_{0},V_{0},N_{0}$, and $\xi_{0}(t)$,
$S_{0\text{in.eq}}(E_{0},V_{0},N_{0},\xi_{0}(t))$ is its maximum possible
value. This non-equilibrium state of the system with the entropy given by
$S_{0\text{in.eq}}$ is said to be in \emph{internal equilibrium}
\cite{Guj-NE-I,Guj-NE-II}, introduced in Sect. \ref{Sect_Second_Law}.

The above discussion can be easily extended to an open system. It has already
been shown \cite{Gujrati-Symmetry} that the Gibbs entropy for the open system
is given by exactly the same formula as Eq. (\ref{Gibbs_Entropy}) except that
$i$ now represents one of the possible microstates of the open system $\Sigma
$. Thus, everything said above applies to the entropy $S(t)$ of $\Sigma$\ by
removing the suffix $0$ above. The only difference is that $E_{0},V_{0},N_{0}%
$, etc. will be replaced by the instantaneous values $E(t),V(t),N.$ The
entropy $S_{\text{in.eq}}$ is a function of $E(t),V(t),\xi(t),N_{0}$,~etc. but
it again cannot have an explicit dependence on time for the same reason that
$S_{\text{in.eq}}$ is already maximum for \emph{fixed }$E(t),V(t),\xi
(t),N_{0}$,~etc. In the discussion below, we only consider the system $\Sigma$.

\subsection{Importance of Equiprobable Microstates for Measurements and
Microstate Entropy\label{Sect_Measurement}}

The equiprobability assumption implies that the system exhibits no bias for
any particular microstate, a point already emphasized by several authors in
the past including Tolman \cite[see Sect. 25, particularly, pp. 63-64]%
{Tolman}, who uses this property of a statistical system as a postulate, when
he discusses the validity of statistical mechanics. This postulate should be
valid even for non-equilibrium states that appear in a system as we vary
macroscopic conditions. This is the main idea about the internal equilibrium
in our approach. The equiprobable or unbiased sampling assumption for the
application of the two Boltzmann probabilities has a very important
consequence for measurements in that one does not have to wait for the system
to sample all of the relevant microstates. The latter is known to take
astronomically large Poincar\'{e} recurrence time \cite{Guj-Recurrence}, as
can be found in any decent textbook on statistical mechanics; see for example,
Huang \cite{Huang}. Let us consider a non-equilibrium system in internal
equilibrium. Because of the equiprobable assumption,
\[
p_{i}(t)=1/W(t),\ \forall i,
\]
where $W(t)$\ denotes the number of microstates in the macrostate at that
instant, so that the average of any thermodynamic quantity like the energy or
entropy is given by%
\[
E(t)\equiv%
{\textstyle\sum\limits_{i}}
E_{i}/W(t),\ S(t)\equiv-%
{\textstyle\sum\limits_{i}}
\eta/W(t)=-\eta,
\]
in which the sum is over $W(t)$ microstates. In reality, equiprobable
microstates do not have to imply an exact equiprobability; they can be within
statistical error. One can think of $-\eta\equiv\ln W(t)$ as the entropy per
microstate or the entropy of a microstate \emph{under the assumption of
equiprobability}. As the observables in each of these microstates take values
within statistical fluctuations of the average observables, even a few samples
will result in a highly reliable value of the observables. The only difference
is that we need to replace $W(t)$ by the number of samples. This is what makes
classical thermodynamics so reproducible within statistical fluctuations. For
example, let us take a \emph{single sample}, which happens to be in some
microstate of energy $E$ with probability $1/W(t)$. The value of $E$ is within
statistical error to the average energy $E(t)$. There is no sum in the
definition for $E(t)$ now. The result is that%
\begin{equation}
E(t)=E,\ \ S(t)\equiv-\eta=\ln W(t).\label{Microstate_Entropy}%
\end{equation}
A single sample, or equivalently a single microstate with probability
$p(t)=1/W(t)$, provides us with the energy $E$ within statistical error and
with the Boltzmann entropy. The latter is not zero and contradicts the UCA
Conjecture \ref{Marker_Microstate}. The same is also true of other
observables. 

\begin{remark}
\emph{There is no need to take the average over a large ensemble or over a
long period of time.} 
\end{remark}

This is why only a few samples to obtain average thermodynamic quantities give
rise to highly reproducible results in thermodynamics. One most certainly does
not have to take a very long time average or a very large ensemble average.
The above discussion shows how the measurement will not affect the
thermodynamic properties, in particular, the entropy of the system in
accordance with the expectation noted in Sect. \ref{Sect_Statistical_Entropy}.
The requirement that the measurement should have ample time to \emph{sample
all relevant microstates} $W(t)$ is not only unnecessary but also not physical
as that time is comparable to the Poincar\'{e} cycle
\cite{Guj-Recurrence,Gujrati-Symmetry,GujratiResidualentropy}. We believe that
GMc have unnecessarily confused the issue by their following suggestion
\cite{Mauro}:

\begin{quotation}
\textit{Consequently, only the time average can correctly reproduce the
measured properties of glass. The underlying reason for this is that at any
instant in time a system has one and only one representative point in phase
space. The properties measured during a given experiment are a result of
averaging over only those microstates that are accessed by the system during
the measurement time. This, in a nutshell, is the principle of causality. For
short observation times, only a small number of microstates are accessed. For
long observation times, a large number of microstates are accessed}.
\end{quotation}

Of course, it is possible in some rare cases that the sample we have is not a
representative of internal equilibrium. In that case, we will obtain results
that are not reproducible. But such a situation will be truly rare.

\subsection{Gibbs vs Boltzmann Entropy Formulation}

It should be clear form above that the Gibbs formulation is more general than
the Boltzmann formulation
\cite{Guj-Irreversibility,Guj-Recurrence,Gujrati-Symmetry,GujratiResidualentropy}%
. In both cases considered above, the Boltzmann entropy is the maximum
possible entropy which occurs only when the \emph{available microstates have
become equiprobable}, and the system is either in equilibrium or in internal
equilibrium \cite{Guj-NE-I,Guj-NE-II}. The system is said to be in equilibrium
when Eq. (\ref{Boltzmann_Entropy}) determines the entropy, and in internal
equilibrium when Eq. (\ref{Boltzmann_Entropy(t)}) determines the entropy. In
all other cases, the Gibbs entropy is the correct entropy of the system. As
the Gibbs formulation supersedes the Boltzmann formulation, it seems to be the
more general one to use for non-equilibrium systems. The relevance of the
Gibbs formulation of entropy for non-equilibrium systems has been discussed
recently
\cite{Guj-Irreversibility,Guj-Recurrence,Gujrati-Symmetry,GujratiResidualentropy}
by us, and we refer the reader to them for more details. We should, however,
mention that Boltzmann's \emph{H}-theorm already shows that the Gibbs
formulation is more general and conforms to the second law. We have also
discussed \cite{Gujrati-Symmetry,GujratiResidualentropy} there how the
time-average is not very useful at low temperatures.

With the above discussion of the statistical entropy, we now turn to UCA. GMc
\cite{Mauro} go on to state as a fact that "\ldots\textit{the Gibbs entropy is
valid for canonical systems in equilibrium,\ldots cannot be used in
nonequilibrium systems since it implicitly assumes ergodicity.}" This is far
from the truth; see above also. The Gibbs entropy is valid for any system
(isolated or not), which need not be in equilibrium. We refer the reader to
the derivation of the Gibbs entropy formulation in Eq. (40.7) for a
non-equilibrium ideal gas in the famous textbook by Landau and Lifshitz
\cite{Landau}; when this entropy is maximized, it gives the \emph{grand
canonical distribution}. But the point is that the Gibbs entropy is valid even
for non-equilibrium systems. It also does not require ergodicity. It should be
stressed that Gibbs never mentions ergodicity in his famous treatise
\cite{Gibbs}. The entropy of a non-equilibrium isolated system using Gibbs
formulation is considered by Tolman \cite{Tolman} to show that it is a
non-decreasing function of time and satisfies the second law. Using this
formulation for the isolated system, it is easy to show
\cite{Gujrati-Symmetry,GujratiResidualentropy,Guj-NE-II} that the same
formulation also applies to open systems. GMc use their above limited view of
the Gibbs entropy to argue that the approach by Lebowitz and Goldstein of
using the Boltzmann entropy formulation "\textit{\ldots is the only one valid
and consistent with the Second Law of nonequilibrium thermodynamics;}" see
Claim \ref{Marker_Boltzmann}. This is a very strong statement with the
implication that it is the truth to be accepted by the reader. Unfortunately,
the statement is not the truth as Gibbs formulation of the entropy also
satisfies the second law as we have discussed above. Moreover, it is also not
accepted by all workers in non-equilibrium thermodynamics. Even Ruelle
\cite{Ruelle}, who is cited by GMc \cite{Mauro}, categorically
\emph{disagrees} with the interpretation in UCA We quote Ruelle \cite{Ruelle}

\begin{quote}
"\textit{The fact that we take seriously the expression }$S(\rho)=-\int
dx\rho(x)\ln\rho(x)$\textit{ for the entropy seems to be at variance with the
point of view defended by Lebowitz,}$^{\text{(20)}}$\textit{ who prefers to
give physical meaning to a \emph{Boltzmann entropy} different from }$S(\rho
)$\textit{. There is, however, no necessary contradiction between the two
points of view, which correspond to idealizations of different physical
situations. Specifically, Lebowitz discusses the entropy of states which are
locally close to equilibrium, while here we analyze entropy production for
certain particular steady states (which maybe far from equilibrium)."}
\end{quote}

It is our opinion that GMc have unnecessarily confused the issue of the
statistical interpretation of the entropy. While they argue for the
superiority of the Boltzmann entropy for which no reasonable arguments are
offered, they go back to use the Gibbs formulation, which they blame to be an
equilibrium quantity, knowing well that the glass is not an equilibrium
system. We find nothing wrong with the Gibbs formulation, contrary to the
implications in UCA.

The suggestion by GMc that the glass confined to a component is like a
canonical system at fixed temperature and volume (while it is really a
non-canonical system with time-varying temperature and constant pressure)
misses out the most important aspect of the glass transition. The temperature
controlling the vibrations within the component and the temperature describing
component hopping over a longer period of time are two distinct temperatures.
As they do not include any internal variable in their approach, they miss out
in capturing all non-equilibrium contributions to the problem. All they seem
to be concerned with is to justify the loss of entropy using a computational
approach. Proposing a computational approach that shows entropy of the glass
below that of the supercooled liquid is not a proof of the conjecture of the
entropy loss. We need to ensure that the resulting physics is consistent with
the established laws of physics, such as the second law. We now turn to this
aspect of their approach.

\subsection{Ergodicity and Causality}

When the entropy is given by Eq. (\ref{Boltzmann_Entropy}), the system is said
to be \emph{ergodic}. A system is either ergodic or it is not. When the
entropy is given by Eq. (\ref{Boltzmann_Entropy(t)}), one can say that the
system is "ergodic with respect to the available microstates belonging to
$W(t)$." But this is not equivalent to the original concept of ergodicity,
which is mathematically defined \cite{Gallavotti,Patrascioiu,Szasz} by
requiring the equality of infinite time and ensemble or phase-space averages;
see also Tolman \cite{Tolman}:%
\[
\overline{f}=\left\langle f\right\rangle ,
\]
where
\[
\overline{f}\equiv\lim_{t\rightarrow\infty}\frac{1}{t}\int_{0}^{t}f(t^{\prime
})dt^{\prime},\ \ \,\left\langle f\right\rangle \equiv\int_{\Gamma
}f(p,q)dpdq/\int_{\Gamma}dpdq.
\]
The infinite-time average is required to ensure that the average does not
depend on the initial state of the system. Thus, the Deborah number%
\[
D_{\text{e}}(t,\tau_{\text{relax}})\equiv\frac{\tau_{\text{relax}}}{t}%
\overset{t\rightarrow\infty}{\rightarrow}0
\]
if we wish to test whether a given system is ergodic or not. If we observe a
system over a period much shorter than $\tau_{\text{relax}}$, so that
$D_{\text{e}}>>1$, all we observe is a non-equilibrium system, but it tells us
nothing about the system being ergodic or not. That can only be answered by
observing a system much much longer than $\tau_{\text{relax}}$; indeed this
time must be comparable to the Poincar\'{e} cycle. Even if we observe the
system for a period $t$ comparable to $\tau_{\text{relax}}$, the system has
not have enough time to visit all relevant microstate. In this case,%
\[
\overline{f}(t\sim\tau_{\text{relax}})\equiv\frac{1}{t}\int_{0}^{t}%
f(t^{\prime})dt^{\prime}%
\]
will be dominated by microstates that the system has visited during $t\sim
\tau_{\text{relax}}$; but these microstates have a strong correlation with the
initial state, which may be far from equilibrium. Thus, such a finite
time-average will not be equal to the ensemble average $\left\langle
f\right\rangle $ for the system, and one would incorrectly conclude that the
system is non-ergodic, even if it is ergodic. It should be clear that because
of the limit $t\rightarrow\infty$, ergodicity is a property of an equilibrium
state. It has no meaning for a non-equilibrium state. Therefore, any
suggestion that a glass is non-ergodic requires the \emph{phenomenological
assumption} that it is a permanently frozen structure. This is most certainly
not a valid assumption in the glass transition region. Thus, to speak of
ergodicity breaking at or near $T_{0\text{g}}$ is a misnomer in our opinion,
even though it is loosely used in the literature.

In practical terms, a system is \textquotedblleft ergodic\textquotedblright%
\ if, after \emph{sufficiently long} time $t>>\tau_{\text{relax}}$, it visits
all possible microstates consistent with a macrostate with \emph{equal
probability}. This is no different than the above mathematical definition, as
the time required to visit all microstates is comparable to the Poincar\'{e}
cycle \cite{Guj-Recurrence,Gujrati-Symmetry,GujratiResidualentropy}. However,
the practical definition, which uses the macrostate, causes the following
problem. It depends on how the macrostate is defined. As we have seen in Sect.
\ref{Sect_Second_Law}, the concept of a macrostate in non-equilibrium systems
depends on time and will eventually become the equilibrium macrostate when
$t\approx\tau_{\text{relax}}$. This will make every system ergodic, whether
the equilibrium state is unique or not, such as an Ising ferromagnet which has
two distinct equilibrium states, and for which the macrostate can be described
by the magnetization along with other observables. If, however, the
magnetization is not used in specifying the macrostate, then the practical
definition will show that the ergodicity is broken in ferromagnets. Usually,
we require the equilibrium state to be not unique for the loss of ergodicity.
Therefore, we believe that the mathematical definition of ergodicity as a
limiting property is the proper way to investigate ergodicity. Such a
definition will surely make the liquid above the melting temperature ergodic.
Now, just because we observe an ergodic system such as this liquid at some
time $t<\tau_{\text{relax}}$, so that we observe a non-equilibrium state of
the system, does \emph{not} make the ergodic system non-ergodic. The system
remains ergodic as it will eventually equilibrate to its unique equilibrium
state for $t\approx\tau_{\text{relax}}$ in accordance with its ergodic nature.
Therefore, to speak of ergodicity restoration for a glass is not proper as the
glass fully relaxes. From the proper mathematical definition of ergodicity, a
glass is also "ergodic" as it will eventually equilibrate to the unique
equilibrated supercooled liquid. The quotation marks on ergodic here is to
reflect the fact that we are taking the corresponding crystal out of the
consideration. These issues have been discussed elsewhere
\cite{Gujrati-Symmetry,GujratiResidualentropy}.

Palmer \cite{Palmer} does talk about the loss of ergodicity, but it is
understood that the relevant part of the phase space is a union of
\emph{disjoint} components with no possible transitions between them; the
union of these components determine the macrostate. However, the system will
be confined\emph{ forever} to one of these components, so that we can set
$\tau_{\text{relax}}\rightarrow\infty$ and $D_{\text{e}}\rightarrow\infty$.
The situation is similar to that in a ferromagnet, except that there are many
more macrostates considered by Palmer. The ergodicity is clearly lost in this
case. However, the situation with glasses is quite different since
$\tau_{\text{relax}}<\infty$ so that $D_{\text{e}}\rightarrow0$ in the limit.

We believe that GMc unnecessarily complicate the issue of glass transition by
invoking the loss of ergodicity just because we happen to observe the system
in its non-equilibrium state at time intervals $t<\tau_{\text{relax}}$. If we
accept this rendition of ergodicity loss, even a liquid above its melting
temperature will become non-ergodic at $t<\tau_{\text{relax}}$, and no useful
purpose is served by introducing such a concept of ergodicity. We refer the
reader to a very good discussion of ergodicity by Tolman \cite{Tolman} and by
Gallavotti \cite[see, in particular, p. 257]{Gallavotti}. In our view, the
glass transition at $T_{0\text{g}}$ is a transition from equilibrated
supercooled liquid to a non-equilibrium supercooled liquid and the transition
at $T_{0\text{G}}$ a transition from this non-equilibrium supercooled liquid
to a glass, which is almost solid and its structure appears frozen over a long
period of time ($t>>\tau_{\text{obs}}$). Thus, our interpretation is different
from GMc.

GMc also refer to the concept of causality in their work; see Conjecture
\ref{Marker_entropy_loss}. This issue seems to be first raised by Kievelson
and Reiss \cite{Kievelson,Reiss}. It basically refers to the possible
existence of a large number of degenerate microstates for a macrostate at
absolute zero. According to Reiss \cite{Reiss}

\begin{quote}
"\textit{Besides the residual entropy at 0 K being an artifact resulting from
apparent entropy measurements along at least partially irreversible paths,
this specification is incompatible with a view of the second law which
establishes entropy as a function of state. If it is a state function it
depends only on its measured state, not upon the history of the system and
certainly not upon its future. Since the system does not visit its alternative
degenerate states during the time of measurement, it is unaware of these
states, and the principle of causality forbids it to be affected by these
states.}"
\end{quote}

The entropy in non-equilibrium thermodynamics is a \emph{generalized} state
function in that it is not only a function of instantaneous observables but
also internal variables. The internal variables are no longer independent of
the observables only when the system has come to equilibrium. Only the
observables, and not the internal variables, are measurable. Thus, Reiss
proposes a very narrow concept of entropy used in non-equilibrium
thermodynamics. As the instantaneous observables and internal variables
clearly depend on the history, Reiss's assessment about the history dependence
is incorrect. Moreover, as the second law destroys time-reversibility, the
system is very much controlled by its "future," i.e. the equilibrium state.
Every system, no matter how it is prepared, "knows" exactly where its future
lies and relaxes towards it. The idea of causality in the above quote with
respect to the statistical entropy is somewhat misleading. Just as each role
of a die results in an independent outcome, yet their probabilities are not
independent (after all, they have to add to unity), different microstates are
independent, yet their probabilities are not
\cite{Gujrati-Symmetry,GujratiResidualentropy}. Let us clarify this by a
simple example discussed by us elsewhere
\cite{Gujrati-Symmetry,GujratiResidualentropy}. Let us throw an six-face
unloaded die. Let the outcome of the throw be $5$. Then, we have
\[
p_{i}=\delta_{i,5},\ \ \ i=1,2,\cdots,6,
\]
where $\delta_{i,j}$ is the kronecker delta, and where $i,j$ denote the six
possible outcomes. Let us assume that the outcome of the next throw is $3$.
Then, $p_{5}=p_{3}=1/2$, and all other probabilities remaiin zero. Even if the
two throws are independent, the probaility distribution changes depending not
only on the numeber of throws, but also on the particular outcomes. As the
entropy is determined by the probability distribution, it should not come as
surprise that the microstates (throws here), though independent, affect the
value of the entropy. Only when the number of throws has become so large that
$p_{i}\rightarrow1/6$, the "equilibrium value," can we say that additional
throws will not affect the entropy. But this is precisely the property of an
equilibrium state.

It should be clear that the probabilities of independent events are not
independent in probability theory. As entropy is a statistical quantity (after
all it is the average of the negative of the index $\eta$), its value is
determined by microstate probabilities. Therefore, the entropy is a measure of
the index of probability of all "independent" microstates. Causality has
nothing to do with the concept of statistical entropy. Let us consider the
case when microstates are equiprobable. The entropy of any sample at $t$ (a
single microstate at that instance) is given by its probability, as shown in
Eq. (\ref{Microstate_Entropy}). Mechanical quantities such as energy, volume,
etc. are not affected by this probability. The mechanical quantities are
independent for each sample, as expected. But entropy, being a statistical or
thermodynamic quantity, is determined by the probabilities, which themselves
are controlled by the sum rule%
\[
\sum_{i}p_{i}(t)\equiv1
\]
over all microstates or samples, so that the probability is \emph{determined}
by the number of microstates $W(t)$. This expected result has nothing to do
with the temperature such as the absolute zero and remains valid at all
temperatures and at all times whenever internal equilibrium is present.

\section{Loss of Entropy in UCA and the Glass
transition\label{Sect_Entropy_Loss}}

Continuous vitrification results in the entropy curves (thermodynamic and
statistical entropy $S(T_{0},t)$ of Glass1 in CA and statistical entropy
$\widehat{S}(T_{0},t)$ of Glass2 in UCA) in Fig. \ref{Fig_entropyglass}. Let
us consider our system above $T_{0\text{g}}$, where the system is either the
equilibrated supercooled liquid or the equilibrated liquid. Let $E^{\prime
},V^{\prime}$ and $S^{\prime}$ denote the energy, volume, and entropy of the
equilibrated state at this temperature $T_{0}^{\prime}$, respectively. The
equilibrium value of the internal variable is denoted by $\xi^{\prime}$. At
time $t=0$, we abruptly bring this system in \emph{contact} with another
medium at temperature $T_{0}$. Immediately prior to the instant the contact is
made, the system is in some microstate $i^{\prime}$, but we do not know
precisely which microstate it is in. There is a probability $p_{i^{\prime}}$
that the system is in microstate $i^{\prime}$. Let $\tau$ denote the time
required for $i^{\prime}$ to evolve to another microstate at $T_{0}$. This
microstate has no time to change immediately after the contact, so the system
remains in the same microstate initially for $t<\tau$.

What is the entropy $S(T_{0},t)$ for $t<\tau$ after the contact?

\subsection{The Unconventional Approach}

According to Conjecture \ref{Marker_Boltzmann}, the statistical entropy
$\widehat{S}(T_{0},t)$ is identically \emph{zero} (recall that we are
considering the entropy and not just the configurational entropy here and in
Fig. \ref{Fig_entropyglass}; similarly, our microstate refers to the system
and not just to its configurational state):%
\[
\widehat{S}(T_{0},t)\equiv0\ \ \ \text{for }t<\tau\text{.}%
\]
This will be true regardless of whether $T_{0}>T_{0}^{\prime}$ or $T_{0}%
<T_{0}^{\prime}$. We have already used this argument in Sect. \ref{Sect_UCA},
which we will now formalize. This zero entropy for $t<\tau$ will result in an
entropy curve similar to the entropy curve of Glass2 in Fig.
\ref{Fig_entropyglass} at $T_{0}$\ in that it lies \emph{below} $S_{\text{eq}%
}(T_{0})$, except that it is identically zero for $t<\tau$. The argument works
the same way even if $T_{0}^{\prime}$ and $T_{0}$ happen to be above the
melting temperature $T_{\text{M}}$, where we have an ordinary liquid, which is
not thought to lose ergodicity.

\begin{conclusion}
It thus follows that the argument of the entropy loss in UCA has nothing to do
with any impending glass transition or any impending loss of ergodicity, both
of which require temperatures near $T_{0\text{G}}$. It is merely a consequence
of two distinct facts:
\end{conclusion}

\begin{enumerate}
\item[(a)] \textit{the duration of observation }$\tau_{\text{obs}}<\tau$ (we
will see below that this restriction on observation time is toally irrelevant
for the conclusion)\textit{; }

\item[(b)] \textit{the entropy of a microstate is zero per Conjecture
\ref{Marker_Microstate}}.
\end{enumerate}

Let us now consider the above thought experiment in time at any temperature
$T_{0}$. The entropy is $\widehat{S}(T_{0},t)\equiv0\ $for $t<\tau$. We now
watch the microstate $i^{\prime}$ to evolve to some other microstate
$i^{\prime(1)}$ at $t=\tau$, and let $\tau^{(1)}$ be the evolution time for
$i^{\prime(1)}$. Since the system is in a single microstate, it follows from
Conjecture \ref{Marker_Microstate} \ that the entropy of the system is still
zero for $t<\tau+\tau^{(1)}$. We wait till $t=\tau+\tau^{(1)}$ so that the
current microstate evolves into another microstate $i^{\prime(2)}$, and let
$\tau^{(2)}$ be the its evolution time. From the same reasoning, we find that
\[
\widehat{S}(T_{0},t)\equiv0\ \ \ \text{for }t<\tau+\tau^{(1)}+\tau^{(2)}.
\]
It is easy to see that
\[
\widehat{S}(T_{0},t)\equiv0\ \ \ \text{for }t\leq\infty.
\]
This makes the second part of Conjecture \ref{Marker_Microstate} inconsistent
with its first part.

\begin{conclusion}
Conjecture \ref{Marker_Microstate} cannot be justified.
\end{conclusion}

Even though we have rejected Conjecture \ref{Marker_Microstate}, let us assume
that the entropy $\widehat{S}(T_{0},t)\ $continues to increase in time from
its initial value $\widehat{S}(T_{0},0)\equiv0$ for reasons not clearly
specified by GMc. We should recall, see
Remark\ \ref{Marker_Remark_GMc_entropy}, that the statistical concept of
entropy adopted by GMc cannot entertain the second law. So, its increase must
be justified on some other grounds, which GMc have not done so far. Within the
framework of the unconventional approach, let us ask: what would happen if
$t=\tau_{\text{obs}}$? If the relaxation time $\tau_{\text{relax}}%
<\tau_{\text{obs}}$, the entropy $\widehat{S}(T_{0},t)$ will continue to
increase and become equal to the equilibrium entropy. For $T_{0\text{g}}%
<T_{0}<T_{\text{M}}$, the entropy will equal $S_{\text{eq}}(T_{0})$ of the
supercooled liquid. For $T_{0}<T_{0\text{g}}$, $\widehat{S}(T_{0},t)$ will
continue to increase from zero and become equal to the entropy $\widehat
{S}(T_{0},\tau_{\text{obs}})$ of Glass2, see the horizontal bars on upward
pointing arrows in Fig. \ref{Fig_entropyglass}, at $t=\tau_{\text{obs}}$ as it
tries to grow to its equilibrium value $S_{\text{eq}}(T_{0})$ for reasons not
mentioned in UCA. If we disrupt the time-evolution at $t=\tau_{\text{obs}}$
such as by abruptly changing the temperature of the medium, the system will
have this entropy $\widehat{S}(T_{0},\tau_{\text{obs}})$ at the moment the
change is made. According to our understanding of UCA, this is the glass
transition in the system. If we let the system relax at $T_{0}$, the entropy
will continue to increase form $\widehat{S}(T_{0},\tau_{\text{obs}}),$ this
time from above the horizontal bar on the upwards arrow, until it reaches
$S_{\text{eq}}(T_{0})$ as $t$ $\rightarrow\tau_{\text{relax}}(T_{0})$. The
entropy is \emph{always increasing}, with the glass transition playing no
special role in the growth of the entropy. We see no evidence of this process
being inverse of the glass transition at $t=\tau_{\text{obs}}$; entropy
$\widehat{S}(T_{0},t)$ is an increasing function of $t$ at all times:

\begin{conclusion}
We see no real difference in the way entropy behaves during the evolution of
the system, which suggests that the glass transition and relaxation are not
inverse processes.  
\end{conclusion}

\subsection{Entropy Loss and the Second Law}

Regardless of the amount of drop (it does not even have to be comparable to
$S_{\text{res}}$), the statistical entropy $\widehat{S}(t)$ of such a
non-equilibrium state in UCA must approach that of the supercooled liquid from
below. This will result in the increase of the entropy during relaxation,
which violates Eq. (\ref{Entropy_variation0}). As this equation was a
consequence of the second law, we come to the following

\begin{conclusion}
A conjecture of entropy drop below that of the supercooled liquid will violate
the second law as is clear from Eqs. (\ref{Total_Entropy_Rate}),
(\ref{Entropy_variation1}) and (\ref{Entropy_variation0}). Thus, the
statistical entropy $\widehat{S}(T_{0},t)$ and the thermodynamic entropy
$S(T_{0},t)$ are two distinct quantities, with the statistical entropy
$\widehat{S}(T_{0},t)$ having no relevance to the glassy state.
\end{conclusion}

\subsection{Our Conventional Approach}

We believe that the core of the problem with UCA is the conjecture about the
entropy of a microstate; see Conjecture \ref{Marker_Microstate}. The entropy
is a property of a macrostate. However, if the system is in internal
equilibrium or in equilibrium, then one can obtain the entropy of the system
by simply using a microstate \cite{Guj-Irreversibility,Guj-Recurrence}, as
seen in Eq. (\ref{Microstate_Entropy}). This entropy is not always zero; it
will be zero if and only if the microstate is unique. The \emph{macrostate}
corresponding to given $E(t),V(t)$ and $\xi(t)$\ is the collection of all
$W(t)$ microstates with given $E(t),V(t)$ and $\xi(t)$ along with their
probabilities \cite{Gujrati-Symmetry}. All instantaneous thermodynamic
averages including the instantaneous entropy are average quantities over the
macrostate at that instant. Under the assumption of internal equilibrium, the
instatntaneous averages can be obtained from a single microstate or sample, as
discussed in Sect. \ref{Sect_Measurement}. The dynamics within a glass for it
to jump from one microstate to another in time is not necessary for
determining these instantaneous averages, an issue discussed elsewhere
\cite{GujratiResidualentropy,Gujrati-Symmetry} to which we refer the reader
for details. When we pick a glass, or when we make an instantaneous
measurement, we do not know which microstate it belongs to. All we know is the
probability $p_{i}$ for the microstate $i$. If the glass formation occurs
under an unbiased condition, all microstates will be equally probable so that%
\[
p_{i}\equiv1/W(t),
\]
and we obtain the Boltzmann entropy $\ln W(t)$. Let $W_{\text{G}}$ denote the
number of possible microstates at absolute zero. When a glass is formed, it is
equally likely to be in any of the $W_{\text{G}}$ microstates at absolute zero
so that the residual entropy resulting from this will be
\[
S_{\text{res}}=\ln W_{\text{G}}.
\]
The residual entropy will be zero if and only if we know for sure that the
glass is a particular microstate, which will happen only if $W_{\text{G}%
}\equiv1.$ This we believe will represent an ideal glass. Just because one
glass sample at absolute zero is in some microstate out of $W_{\text{G}}$
$(>1)$ does not mean that the glass entropy is zero. The latter would be the
case if we knew which particular microstate the glass sample happens to be.
Only when $W_{\text{G}}\equiv1$ can we be sure that all glass samples would be
in the same microstate, and the glass entropy would be precisely zero
\cite{Gujrati-Symmetry}. In all other cases, all we know is that the
probability that the system is in microstate $i$ is $p_{i}$, and the entropy
is given by the Gibbs entropy in Eq. (\ref{Gibbs_Entropy}). 

There is another way to understand this probability
\cite{GujratiResidualentropy,Gujrati-Symmetry}. We consider dividing the
system into a large number of macroscopically large but quasi-independent
parts of equal size, each of which can be in any microstate $\iota$ associated
with a part with a probability $p_{\iota}$. Then the entropy $s(t)$ of each
part is given%
\[
s(t)\equiv-%
{\textstyle\sum\limits_{\iota}}
p_{\iota}(t)\ln p_{\iota}(t),
\]
and the entropy of the system, using its additive property, is given by
\[
S(t)=\sum s(t)\equiv N_{\text{P}}\overline{s}(t),
\]
where the sum is over all $N_{\text{P}}$ parts of the system and $\overline
{s}(t)$ is the average entropy per part.

Once we recognize that the entropy of a microstate is in general not zero
identically, we have no problem understanding that when we bring the system in
contact with a medium at another temperature, the entropy immediately after
the contact is also unchanged. It then changes towards the new equilibrium
value during its relaxation, which is shown by downward arrows in Fig.
\ref{Fig_entropyglass}. If we interrupt this relaxation at $t=\tau
_{\text{obs}}$ ($T_{0}<T_{0\text{g}}$) by bringing the system in contact with
a different medium at a lower temperature, the system will not completely
relax. The current value of the entropy $S(T_{0},t=\tau_{\text{obs}})$ becomes
the initial value of the entropy at the new temperature. A sequence of such
interruptions will eventually result in a "frozen" glass below $T_{0\text{G}}%
$.  This understanding of the microstate entropy also shows that one does not
have to observe the system over a period necessary to sample \emph{many} of or
\emph{all} of the microstates associated with the macrostate or one does not
require that the entropy is maximum only when all the microstates have been
visited. The latter understanding of entropy has been criticized in the past
by several authors; see for example Huang \cite{Huang}, Tolman \cite{Tolman} ,
Gallovatti \cite{Gallavotti}, Gujrati \cite{Guj-Recurrence}, and the argument
has been revisited recently by Goldstein \cite{Goldstein}: the time needed for
\emph{all} the microstates of a macroscopic system to be visited so that the
entropy becomes maximum is beyond the current age of the universe. We have
already argued against the time-average to be relevant for any measurement
\cite{Gujrati-Symmetry,GujratiResidualentropy}. In this work, we have shown
clearly that a single instantaneous measurement is sufficient to provide us
with a thermodynamic description of the system at that instant. Any
measurement that takes some finite non-zero duration will never yield any
instantaneous information about the system.

\section{Discussion and Conclusions}

We have briefly described and extended the conventional approach we have
developed earlier for any non-equilibrium system. We consider the system
$\Sigma$ to be surrounded by a very large medium $\widetilde{\Sigma}$ to form
the combined system $\Sigma_{0}$ so that the fields of the medium are not
affected by the presence of processes going on inside the system, whatever
they may. Thus, the approach can be applied to glasses; some authors sometimes
identify them by some stretch of imagination as non-ergodic. However, whether
ergodicity is lost or not plays no role in the behavior of the collection
$\Sigma_{0},$ which we treat as an isolated body so that its thermodynamic
entropy cannot decrease with time. This is the statement of the second law for
the isolated body. We assume that both the body and the medium are separately
in internal equilibrium, but not in equilibrium with each other. The internal
equilibrium allows us to introduce instantaneous fields $T(t),P(t)$, etc. for
the system and the constant fields $T_{0},P_{0}$, etc. for the medium. We also
use a single internal variable $\xi(t)$, in addition to $S(t),V(t)$ as induced
internal variables, to describe possible relaxation in the system as it
approaches equilibrium. The non-equilibrium nature of the system appears in
the values of instantaneous fields $T(t),P(t)$, etc. being different from
$T_{0},P_{0}$, etc. of the medium. These differences in the fields cause
non-negative irreversible entropy generation in accordance with the second
law. In an isobaric vitrification, which is what we consider in this work, we
assume that $P(t)=P_{0}$ at all times. The irreversible entropy generation
requires $T(t)\geq T_{0}$; the equality occurs when there is equilibrium
between the system and the medium. The instantaneous observables, internal
variables and entropy at time $t$ are described by the microstate $i_{t}$ the
system happens to be in at that instance along with its probability. This
microstate also represents the instantaneous macrostate of the system. The
effect of an instantaneous measurement is to give the values of the
instantaneous observables and the entropy. The measurement does not alter the
instantaneous value of the observables, internal variables, and the entropy.
The entropy above refers to the \emph{thermodynamic entropy} and its
statistical interpretation is obtained by the Gibbs entropy formualtion: The
\emph{statistical entropy} is a statistical average of the negative index of
microstate probabilities, just like the observables are of mechanical
quantities. In our approach (CA), the two entropies behave in identical
fashion. Any attempt to provide the classical entropy with a statistical
interpretation must satisfy two important requirements:

\begin{enumerate}
\item[CA1.] It must decrease during relaxation in an isobaric vitrification process.

\item[CA2.] Its instantaneous value must not be affected by any instantaneous measurement.
\end{enumerate}

In our approach, any non-equilibrium state, such as the one produced by
changes in the medium by changing its temperature, relaxes towards its new
equilibrium. During such a relaxation under isobaric cooling, the entropy,
ehthalpy and the instantaneous temperature decrease towards their respective
equilibrium values. The relaxation is complete when we wait for $t=\tau
_{\text{relax}}$. During the relaxation process, the system will undergo a
glass transition below $T_{0\text{g}}$, if the system is abruptly changed at
$t=\tau_{\text{obs}}<\tau_{\text{relax}}$ by bringing it in contact with a
medium at a lower temperature. The \emph{instantaneous macrostate} of the
system, described in terms of its observables and internal variable, does not
change when the contact is made. This also means that the entropy also does
not change. If the contact is not made, the relaxation will continue to go on.
Thus, the glass transition and relaxation are part of the same relaxation
process in CA. There is nothing inverse about them. The gain in the
thermodynamic entropy of the medium is shown to be bounded from above by the
maximum change $\Delta S_{\text{eq}}(T_{0})$; the latter is the maximum
possible decrease in the thermodynamic entropy of the system. From a careful
analysis, we have concluded that the thermodynamic entropy $S(T_{0},t)$ of the
system, such as Glass1, must always be above $S_{\text{eq}}(T_{0})$; thus, the
thermodynamic entropy must decrease during relaxation; the decrease is a
consequence of the second law.

The conclusion of the thermodynamic approach is summarized in Theorem
\ref{Theorem_Relaxation}: the thermodynamic entropy must decrease with time
during any isothermal relaxation in isobaric vitrification.

UCA developed by GMc as an attempt to describe glass transition in any
material does not fulfill both requirements CA1 and CA2. Not only that, the
glass transition and relaxation are described as inverse processes. Faced with
these discrepancies and several other unconventional and not adequetly
explained aspects of UCA, we have carefully examined it in this work. To test
the validity of their inverse conjecture UCA5, we decided to treat $\Sigma$ as
part of $\Sigma_{0}$. We do this to determine the entropy gain by the medium
to show unambiguously whether the system can lose so much entropy at the glass
transition that it lies below $S_{\text{eq}}$ in Fig. \ref{Fig_entropyglass};
see UCA5.

GMc incorrectly conclude that the use of classical thermodynamics to calculate
the thermodynamic entropy is logically inconsistent (UCA1). Using this
erroneous conclusion they argue that the residual entropy must vanish in
accordance with the third law (UCA2). However, a careful reconsideration shows
that there is nothing wrong in using the classical non-equilibrium
thermodynamics. Indeed, its use clearly establishes that the calorimetric
entropy $S_{\text{expt}}(0)$ at absolute zero is a \emph{lower bound to the
residual entropy}; the former entropy is normally found to be non-negative,
which makes
\[
S_{\text{res}}\geq S_{\text{expt}}(0)>0.
\]
Thus, the primary motivation of GMc to develop their unconventional approach
UCA is based on an incorrect understanding of classical thermodynamics.\ \ \ \ \ \ \ \ 

As discussed in Sect. \ref{Sect_UCA_Conjectures}, UCA is based on a set of
conjectures, some of which are inter-related, left unproven or satisfactorily
justified by GMc. In particular, as summarized in Remark
\ref{Marker_Remark_GMc_entropy}, their statistical formulation $\widehat
{S}(T_{0},t)$ of the entropy, which is based on the zero-entropy conjecture
UCA3 for a microstate, has nothing to do with the thermodynamic entropy
$S(T_{0},t)$ used in the second law in Eq. (\ref{Second_Law}). The entire UCA
is based solely on this statistical notion of entropy and its computation.
This formulation of $\widehat{S}(T_{0},t)$ in UCA has been developed with a
goal to show entropy loss; yet we find that this formalism, in particular the
growth of the statistical entropy with time in Conjecture
\ref{Marker_Microstate}, is\emph{ inconsistent} with UCA3; the latter, if
accepted, only results in $\widehat{S}(T_{0},t)=0$ at all times under all
conditions including all temperatures. \emph{This is most certainly
unphysical.} In our opinion, it is this conjecture that results in the entropy
loss during a glass transition under cooling.

Even if we allow for the entropy to increase from zero due to some unknown
reasons, not offered by GMc, we find that $\widehat{S}(T_{0},t)$ will always
increase towards $S_{\text{eq}}(T_{0})$ of the equilibrated supercooled
liquid. We find no justification that the relaxation and glass formation are
inverse processes. Their statistical entropy $\widehat{S}(T_{0},t)$ increases
in both processes. The increase of entropy scenario would hold at all
temperatures, not just at and below $T_{0\text{g}}$. Thus, the entropy loss
conjecture has nothing to do with any glass transition or any ergodicity loss;
it is merely a consequence of the zero-entropy conjecture UCA3. The increase
of the statistical entropy $\widehat{S}(T_{0},t)$ in UCA contradicts the
decrease of the thermodynamic entropy found in CA. Thus, \emph{the statistical
entropy in UCA cannot represent the thermodynamic entropy of a glass. }As
their computation also shows an increase of their statistical entropy, their
computational scheme is not useful to understand glasses.

The statistical entropy due to Gibbs that is used in CA remains in conformity
with the\ behavior of the thermodynamic entropy. The entropy of a microstate,
see Eq. (\ref{Microstate_Entropy}), is not necessarily zero. Thus, at each
instant of time, the entropy of a system, which happens to be in a microstate,
is not zero in CA. This instantaneous entropy for an isolated system will
always increase, but for an open system such as a glass may \emph{decrease}.
The latter behavior is in accordance with the second law. In both cases, it is
the irreversible entropy generation that can never be negative. It is our
belief that GMc have overlooked this distinction beteen the entropy and
irreversible entropy generation in their approach, which causes them to
incorrectly believe that the entropy must increase during isothermal
relaxation in vitrification.

It is our opinion that GMc have unnecessarily confused the issue of the
statistical concept of entropy by implying that the Gibbs entropy is not
suitable to describe glasses but the Boltzmann entropy is. This is not a
correct conclusion. Both formulations are appropriate, but care must be
exercised to interpret them properly. Let us assume equiprobability of
microstates. The number of relevant microstats $W(t)$ is most certainly\emph{
not }the number of microstates \emph{sampled} by any measurement for a
macrostate in time, as GMc suggest. The time for that is of the order of
Poincar\'{e} cycle and far exceeds the age of the universe. It is really the
number of microstates \emph{available} to the system, as explained earlier,
and even an instantaneous measurement will give the expected value of the
observables. This is what makes thermodynamics so reliable a science. This
interpretation is the same whether we use the Gibbs entropy formulation or the
Boltzmann entropy formulation. There is no difference between them as long as
we deal with internal equilibrium. Their continuous assertion in various
publication that they are different is most probably due to their
misunderstanding, and serves no purpose except to confuse the issue of the
statistical entropy. There is no reason at all to doubt that the thermodynamic
and statistical entropies are different in any way.

It is our opinion that they have also unnecessarily made too much of an issue
of ergodicity loss and of causality in glasses. All one needs to do is to
treat glasses as a non-equilibrium state and to recall that the statistical
entropy is an average of a statistical quantity, the negative index of
probability as discussed in the work. GMc have taken a very simplistic view of
glasses by ignoring internal variables, two-temperature description and the
fact that fluctuation-dissipation theorems used by them \cite{Mauro} fail for glasses.

We finally conclude that their current theoretical and computational attempts
using UCA has no relevance for glasses.

I am thankful to M. Goldstein for introducing me to the work by GMc, and to
G.P. Johari for his useful comments on an earlier version\ of the manuscript.

\end{document}